\documentclass{emulateapj}
\usepackage{lscape, longtable, hyperref, graphicx, subfigure}
\usepackage{lmodern}
\usepackage[T1]{fontenc}

\newcommand{\Msun}{\ifmmode {M_{\odot}}\else${M_{\odot}}$\fi}
\newcommand{\Rsun}{\ifmmode {R_{\odot}}\else${R_{\odot}}$\fi}
\newcommand{\lapprox }{{\lower0.8ex\hbox{$\buildrel <\over\sim$}}}
\newcommand{\gapprox }{{\lower0.8ex\hbox{$\buildrel >\over\sim$}}}
\def\amin{\ifmmode^{\prime}\else$^{\prime}$\fi}
\def\asec{\ifmmode^{\prime\prime}\else$^{\prime\prime}$\fi}

\slugcomment{To appear in ApJ}
\shorttitle{The Factory \& The Beehive}
\shortauthors{Ag{\" u}eros et al.}

\begin{document}

\title{The Factory and The Beehive\\I.\ Rotation Periods For Low-Mass Stars in Praesepe}

\author{Marcel~A.~Ag{\" u}eros\altaffilmark{1}, Kevin~R.~Covey\altaffilmark{2,3,9}, Jenna~J.~Lemonias\altaffilmark{1},  Nicholas M.~Law\altaffilmark{4,10}, Adam~Kraus\altaffilmark{5,9}, Natasha Batalha\altaffilmark{2}, Joshua S.\ Bloom\altaffilmark{6}, S.\ Bradley Cenko\altaffilmark{6}, Mansi M.\ Kasliwal\altaffilmark{7}, Shrinivas R.\ Kulkarni\altaffilmark{7}, Peter E.\ Nugent\altaffilmark{8}, Eran O.\ Ofek\altaffilmark{7,11}, Dovi Poznanski\altaffilmark{6,8,11}, Robert M.\ Quimby\altaffilmark{7}}

\altaffiltext{1}{Columbia University, Department of Astronomy, 550 West 120th Street, New York, NY 10027} 
\altaffiltext{2}{Cornell University, Department of Astronomy, 226 Space Sciences Building, Ithaca, NY 14853}
\altaffiltext{3}{Visiting Researcher, Department of Astronomy, Boston University, 725 Commonwealth Ave, Boston, MA 02215}
\altaffiltext{4}{Dunlap Institute for Astronomy and Astrophysics, University of
Toronto, 50 St.\ George Street, Toronto M5S 3H4, Ontario, Canada}
\altaffiltext{5}{Institute for Astronomy, University of Hawaii, Honolulu, HI 96822, USA}
\altaffiltext{6}{Department of Astronomy, University of California, Berkeley, CA 94720, USA}
\altaffiltext{7}{Cahill Center for Astrophysics, California Institute of Technology, Pasadena, CA, 91125, USA}
\altaffiltext{8}{Computational Cosmology Center, Lawrence Berkeley National Laboratory, 1 Cyclotron Road, Berkeley, CA 94720, USA}
\altaffiltext{9}{Hubble Fellow}
\altaffiltext{10}{Dunlap Fellow}
\altaffiltext{11}{Einstein Fellow}

\begin{abstract}  
Stellar rotation periods measured from single-age populations are critical for investigating how stellar angular momentum content evolves over time, how that evolution depends on mass, and how rotation influences the stellar dynamo and the magnetically heated chromosphere and corona. We report rotation periods for 40 late-K to mid-M stars members of the nearby, rich, intermediate-age ($\sim$600 Myr) open cluster Praesepe. These rotation periods were derived from $\sim$200 observations taken by the Palomar Transient Factory of four cluster fields from 2010 February to May. Our measurements indicate that Praesepe's mass-period relation transitions from a well-defined singular relation to a more scattered distribution of both fast and slow rotators at $\sim$0.6 $\Msun$. The location of this transition is broadly consistent with expectations based on observations of younger clusters and the assumption that stellar-spin down is the dominant mechanism influencing angular momentum evolution at 600 Myr. However, a comparison to data recently published for the Hyades, assumed to be coeval to Praesepe, indicates that the divergence from a singular mass-period relation occurs at different characteristic masses, strengthening the finding that Praesepe is the younger of the two clusters. We also use previously published relations describing the evolution of rotation periods as a function of color and mass to evolve the sample of Praesepe periods in time. Comparing the resulting predictions to periods measured in M35 and NGC 2516 ($\sim$150 Myr) and for kinematically selected young and old field star populations suggests that stellar spin-down may progress more slowly than described by these relations. 
\end{abstract}


\section{Introduction}
In a seminal paper, Andrew Skumanich (1972) showed that stellar rotation decreases over time such that $v_{rot} \propto t^{-0.5}$, as does chromospheric activity, a proxy for magnetic field strength. This relationship between age, rotation, and activity has been a cornerstone of stellar evolution work over the past 40 years, and has generated almost as many questions as applications. Angular momentum loss due to stellar winds is generally thought to be responsible for the Skumanich law, but the exact dependence of $v_{rot}$ on age is not entirely understood, and relies on the assumed stellar magnetic field geometry and degree of core-envelope coupling \citep{kawaler1988, krish1997}. Furthermore, later-type, fully convective stars appear to have longer active lifetimes than their early-type brethren \citep[e.g.,][]{andy08}, indicating that they are capable of generating significant magnetic fields even in the absence of a standard solar-type dynamo \citep{Browning2008}. The lack of a comprehensive theoretical understanding of the age-rotation-activity relation has not prevented the development and use of gyrochronology, however, which is used to determine the ages of field stars based on a presumed age-rotation relation \citep[e.g.,][]{barnes2007, mamajek2008, collier2009, barnes2010}, nor of empirical age-activity relations, which do not always find activity decaying quite as simply as predicted by the Skumanich law \citep[e.g.,][]{feigelson04,pace2004,giampapa2006}.

Mapping out the dependence of stellar rotation and activity on age requires the study of stars ranging in both mass and age. Statistical constraints on the age-rotation-activity relation can be derived from Galactic field stars \citep[e.g.,][]{feigelson04, Covey2008, irwin2011}, but the homogeneous, coeval populations in open clusters provide an ideal environment for studying time-dependent stellar properties. Ideally, rotation periods for large numbers of cluster members could be measured directly from modulations in these stars' light-curves due to the presence of star spots. There are relatively few nearby open clusters, however, and fewer still have had the high quality photometric data needed to characterize their members' rotation in this manner --- in part because of the sheer difficulty involved in systematically monitoring a large number of stars over several months or more. Studies like that of \cite{skumanich72} relied instead on measurements of the rotational Doppler broadening of spectral lines, a technique that has the advantage of needing only one observation. Translating the resulting $v_{rot}$ sin $i$ measurements into $P_{rot}$ involves making assumptions about stellar radii and inclinations, however, neither of which are well constrained.\footnote{A further limitation of the Doppler broadening technique is that it is sensitive only to stars rotating faster than some threshold set by the spectral resolution.} 

Because of these challenges, our view of the age-rotation-activity relation depended until recently on observations of handfuls of stars in the field and in a small number of well-studied clusters, with the Hyades being a particularly key cluster \citep[e.g.,][]{radick1987, jones1996, stauffer1997, terndrup2000}. Largely because of the advent of time-domain surveys, with their emphasis on wide-field, automated, high-cadence observing, it is now possible to monitor stellar rotation on an entirely new scale \citep[e.g.,][]{irwin2007, meibom2009, hartman2010}. The Palomar Transient Factory \citep[PTF;][]{nick2009, rau2009} provides deep, multi-epoch photometry over a wide field-of-view, and our Columbia/Cornell/Caltech PTF (CCCP) survey, one of PTF's Key Projects, is leveraging this capability to measure rotation periods in open clusters of different ages. 

Our first CCCP target, Praesepe,\footnote{Also known as the Beehive Cluster and M44.} 08 40 24 $+$19 41, is a nearby \citep[$\sim$180 pc;][]{vanleeuwen2009}, rich \citep[$\sim$1200 stars;][]{adam2007}, and intermediate-age \citep[$\sim$600 Myr;][]{delorme2011} cluster that shares many characteristics with the Hyades. Until recently, only five rotation periods --- for mid- to late-M dwarfs --- had been measured for Praesepe members \citep{scholz2007}. These periods were often combined with (less sparse) data for high-mass Hyads in order to infer the mass-rotation relation for 600-Myr-old stars \citep[e.g.,][]{irwin2009}. This was particularly unsatisfying as the Hyades and Praesepe were the two oldest clusters with measured rotation periods, and were therefore essential in studying the evolution of the age-rotation relation from ages of a few 100 Myr to the age of the Sun.

Fortunately, the situation has improved significantly in the past year. \citet{delorme2011} surveyed the Hyades and Praesepe as part of the SuperWASP exoplanet-search program. SuperWASP's sensitivity, tuned to discover exoplanets transiting nearby bright stars, enabled the measurement of rotation periods for 52 late-F to late-K/early-M stars in Praesepe. Meanwhile, \cite{scholz2011} added 49 rotation periods (of which 24 are considered very robust) to the \citet{scholz2007} sample, with the bulk of this new sample being of spectral type M3-M5.\footnote{These spectral types are from the \citet{adam2007} catalog; a few \citet{scholz2007} and \citet{delorme2011} stars lack spectral types in this catalog because they are too faint or too bright.} (Thanks to {\it Kepler}, rotation periods have now also been measured in a 1-Gyr-old cluster, NGC 6811; \citet{meibom2011b}.)

We report stellar rotation periods for 40 late-K/early-M Praesepe members derived from our first season of PTF observations. Our campaign produced $\sim$200 distinct observations of four overlapping fields designed to include a large number of Praesepe members identified by \citet{adam2007}. In Section~\ref{observations} we describe our data and in Section~\ref{periods} our period-finding algorithm. We also compare our periods with those derived by \citet{scholz2007}, \citet{delorme2011}, and \citet{scholz2011} for the stars for which they also measured $P_{rot}$, and flag potential binary systems among our rotators. In Section~\ref{results}, we combine our Praesepe data with that of \citet{scholz2007}, \cite{delorme2011}, and \citet{scholz2011}, and compare color-period relations and mass-period distributions derived from these data to those derived from the Hyades \citep[using data from][]{delorme2011}, the 150-Myr-old clusters M35 and NGC 2516 (\citet{meibom2009} and \citet{irwin2007}, respectively), and kinematically selected young and old field star populations \citep{kiraga2007}, as well as to gyrochrones derived from the models of \citet{barnes2010}. We conclude in Section~\ref{concl}. The Appendix lists interesting variable field stars identified in our Praesepe observations.

In a forthcoming companion paper we use the results of our spectroscopic campaign with the 2.4-m Hiltner telescope at MDM Observatory and the WIYN 3.5-m telescope at NOAO, both on Kitt Peak, AZ, to examine the relationship between rotation and activity in Praesepe.

\begin{figure}[h]
\centerline{\includegraphics[angle=90,width=\columnwidth]{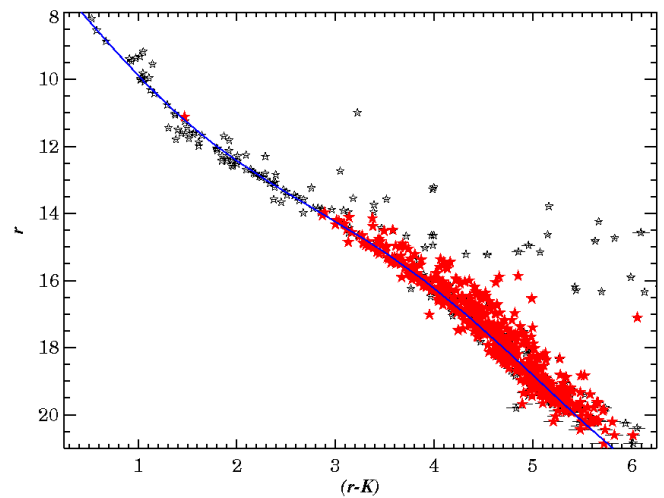}}
\caption{CMD of Praesepe members with 8~$< r_{\rm SDSS} <$~21 mag. Photometric errors are plotted but are typically smaller than the symbols. Overplotted in red are members with PTF detections; the solid line is the cluster main sequence. Stars with $r \sim 14 - 16$ and $(r-K)\ \gapprox\ 4$ are ones for which the SDSS photometry is saturated, and for which accurate photometry could not be calculated from the wings of the point spread function.} 
\label{detections} 
\end{figure} 

\section{Observations and Data Reduction}\label{observations}
The Palomar Transient Factory is a transient detection system comprised of a wide-field survey camera mounted on the automated Samuel Oschin 48 inch telescope at Palomar Observatory, CA, (known as the P48), an automated real-time data reduction pipeline, a dedicated photometric follow-up telescope (the automated Palomar 60 inch), and an archive of all detected sources. The technical aspects of PTF and the project's science goals are described in detail in \citet{nick2009} and \citet{rau2009}. The P48 survey camera is based on the CFH12K mosaic camera formerly at the Canada-France-Hawaii Telescope \citep{rahmer2008}. The camera has 12 chips (one of which is not working), 101 megapixels, 1$\arcsec$ sampling, and a 7.26 deg$^2$ field-of-view. Observations are performed in either Mould $R$ or Sloan Digital Sky Survey (SDSS) $g$, or with a set of H$\alpha$ filters. Under median seeing conditions (1.1$\arcsec$) the camera achieves 2.0$\arcsec$ full-width half-maximum images, and reaches $5$ standard deviation ($\sigma$) limiting AB magnitudes of $m_{g} \approx\ 21.3$ and $m_R \approx\ 21.0$ mag in 60 s exposures \citep{nick2010}. As of 2011 May, the PTF footprint included $7000$ deg$^2$ that have been visited at least 25 times, with nearly $1000$ deg$^2$ having been imaged at least 100 times.

Four overlapping 3.5 $\times$ 2.31 deg fields covering the center of Praesepe were imaged by PTF beginning on 2010 February 2 and ending on 2010 May 19.\footnote{The full extent of our observational footprint was about 18 deg$^2$; see Figure 1 in \citet{covey2010}.}  Because we shared some of our observing time with PTF's transiting-planet search \citep[see][]{law2011}, there were multiple nights early in our campaign and one in April when the cluster was observed every 15 min, resulting in 15-30 images per night. For most of the campaign, the fields were observed 1-2 times a night when the weather allowed, resulting in close to 200 observations for each field (see Table~\ref{obs}). This observing cadence was sensitive to $P_{rot}$ from a few to a few hundred hours, covering the range occupied by the few cluster members with measured periods then known \citep[][]{scholz2007}. 

\begin{deluxetable}{lcc}[!t]
\tablewidth{0pt}
\tabletypesize{\scriptsize}
\tablecaption{PTF Observations of Praesepe \label{obs}}
\tablehead{
\colhead{Field}  & \colhead{Field}   & \colhead{Number of}   \\       
\colhead{Number} & \colhead{Center} & \colhead{Observations}  
}
\startdata
110001 & 08 39 $+$19 15 & 185 \\
110002 & 08 39 $+$20 15 & 189 \\
110003 & 08 44 $+$19 15 & 195 \\
110004 & 08 44 $+$20 15 & 198  
\enddata
\end{deluxetable}

\citet{adam2007} combined data from SDSS \citep{york00}, the Two Micron All Sky Survey \citep[2MASS;][]{2mass}, and USNO-B1.0 \citep{monet2003} to calculate proper motions and photometry for several million sources within 7 deg of Praesepe's center. This census covers a larger area of sky and is deeper than any previous proper motion study of the cluster. The resulting catalog includes 1129 candidate members with membership probability $>$50\% (hereafter referred to as the P50 stars); 442 were identified as high-probability candidates for the first time. \citet{adam2007} estimated that their survey is $>$90\% complete across a wide range of spectral types, from F0 to M5.\footnote{$\sim$40 known members are not included in the \citet{adam2007} catalog as they are too bright for this analysis.}

\begin{figure}[t]
\centerline{\includegraphics[width=\columnwidth]{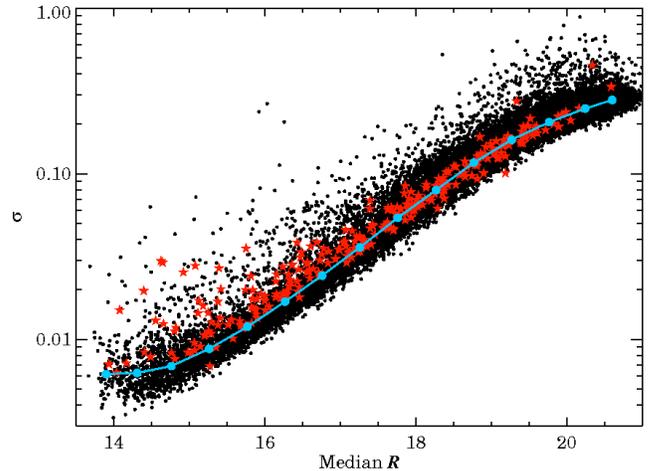}}
\caption{$\sigma$ versus median $R$ magnitude of objects detected in multiple epochs in field 110003. Praesepe members are shown as red stars; the $\sim$30,500 non-members with PTF light-curves in this field are shown in black. At the bright end, the scatter in the light-curves exceeds the formal photometric errors by factors of a few, indicating that the precision is limited by systematic effects rather than by random photometric error. We plot in blue the median $\sigma$ for non-members obtained when placing the magnitudes in bins of width 0.5 mag (e.g., for the 1266 non-members with $16.5 \leq R < 17$ mag, the median $\sigma = 0.02$ mag). Praesepe members have systematically higher $\sigma$ (and are thus more variable) than the median field star for $R\ \lapprox\ 17$.}
\label{std_dev} 
\end{figure}

\begin{figure}[h]
\centerline{\includegraphics[angle=90,width=\columnwidth]{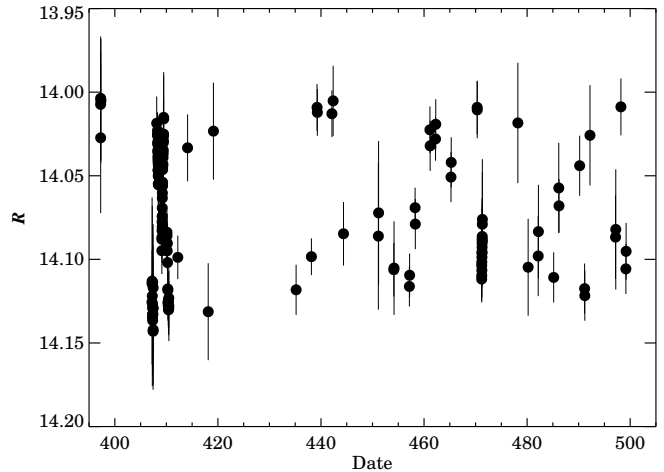}}
\caption{Sample light-curve for the Praesepe member AD 1508 illustrating the mix of observing cadences. The x-axis is the number of Julian days since 2009 January 1; periods of high-cadence observing are visible around days 410 and 470. 142 separate photometric measurements are plotted; the errors are $\sim$0.02 mag on average.}
\label{sample_lc} 
\end{figure} 

Of the 1129 P50 members in the \citet{adam2007} catalog, 923, or over 80\%, lie within the CCCP footprint. Of these, 661 are fainter than the PTF saturation limit ($\sim$14 mag): PTF detected 534 (or 81\%) of these candidate members, with the rest falling within chip gaps or on the dead chip. Figure~\ref{detections} is a color-magnitude diagram (CMD) of Praesepe members derived from SDSS and 2MASS photometry for stars with 8 $< r <$ 21 mag. For plotting purposes we apply minimal quality cuts, requiring only that that the errors in both $r$ and $K$ be $<$0.1 mag. 1105 stars in the \citet{adam2007} Praesepe catalog and 529 of the members detected by PTF meet these criteria. \citet{adam2007} provide spectral types for Praesepe stars based on spectral-energy-distribution (SED) fitting. We apply a polynomial fit to these SEDs and thereby define an interpolated single-star main sequence for the cluster. The PTF-detected members are late-K through early-M stars, as expected given the distance to Praesepe and the PTF exposure time. 

Aperture photometry was measured for each candidate member at each epoch using SExtractor \citep{bertin96}. Precision differential photometry was performed using the pipeline described in \citet{law2011}, and the base zero-points were defined using the SDSS magnitudes of several hundred reference stars. The typical long-term photometric stability is approximately 5 millimag for the brightest stars, and is photon-limited for all stars fainter than $R \sim 15$ mag. Figure~\ref{std_dev} shows the $\sigma$ versus median $R$ for objects detected in multiple epochs for one of the Praesepe fields.  

Positional matching was used to merge detections across epochs, producing a single light-curve for each source. Our Praesepe fields overlapped substantially in the cluster center; for stars observed in multiple fields, light-curves were merged after applying small offsets to the light-curves from each field to ensure a common median magnitude. A sample light-curve for a Praesepe member is shown in Figure~\ref{sample_lc}. 

\section{Period Measurements}\label{periods}
We used a modified version of the Lomb-Scargle algorithm to search our light-curves for periodic signals. We followed Eq.\ 11 of \citet{frescura2008} to define frequency grids based on the number of measurements obtained for each star and the time-span of each light-curve. We oversampled by a factor of five to ensure maximum sensitivity to periodic variability. Lomb-Scargle periodograms were then computed iteratively: an initial periodogram was calculated from those data points within $6\sigma$ of the light-curve's median magnitude. The light-curve was phase-folded using the period corresponding to the frequency with the maximum periodogram power and smoothed with a boxcar window whose width was equivalent to 10\% of the data points within the full light-curve.  

Residuals were calculated for each point with respect to this phased, smoothed light-curve. Points with residuals greater than $4\sigma$ from the smoothed, phased light-curve were rejected before calculating a new Lomb-Scargle periodogram. After two iterations, the frequency with the maximum power was selected as the most likely period for that star. Potential beat frequencies between the primary periodogram peak and a possible one-day alias, typical for ground-based, nightly observing campaigns, were flagged following Eq.\ 1 of \citet{messina2010}. In this manner we measured potential rotation periods for all of the cluster members detected by PTF.  

To test the significance of these periods, we performed a permutation test on our light-curves \citep[][]{efron1982}.\footnote{For a similar approach, but using randomly generated magnitudes, see \citet{frescura2008}.} We conducted the analysis described above on each light-curve after randomly scrambling the magnitudes measured at each epoch. Repeating this test 100 times on each scrambled light-curve, we identify the maximum measured periodogram peak as the power threshold corresponding to a $<$1\% false alarm probability (FAP) in the absence of ordered variations. Across our entire sample, this analysis established that a periodogram peak with power $\gapprox$25 corresponded to a FAP $<$1\%; indeed, for only three of the 534 stars analyzed here did the 1\% FAP correspond to a periodogram power threshold $>$20. We therefore adopted a conservative power threshold of 30 to select potentially periodic cluster members.

We then visually inspected the output of our search for each candidate. Periodograms were checked to confirm the presence of a single narrow peak, well separated from the underlying background power; further scrutiny established that the periodic behavior was visible and stable across the full light-curve, well-sampled in phase, and of an amplitude at least comparable to the observational noise. The periodograms and phased light-curves for the high-confidence rotators are shown in Figure~\ref{image} and listed in Table~\ref{hicp}. Our analysis produced high-confidence measurements of $P_{rot}$ ranging from 0.52 to 35.85 d for a total of 40 stars. 37 of these stars have $P_{mem} > 95\%$, as calculated by \citet{adam2007}, with two of the other stars having $P_{mem} > 94\%$. The remaining rotator, JS 634, is a relatively low-probability member of the cluster, with $P_{mem} = 62.3\%$. Radial velocity observations will be required to confirm its membership.

\begin{figure}[h!]
\centerline{\includegraphics[angle=-270,width=\columnwidth]{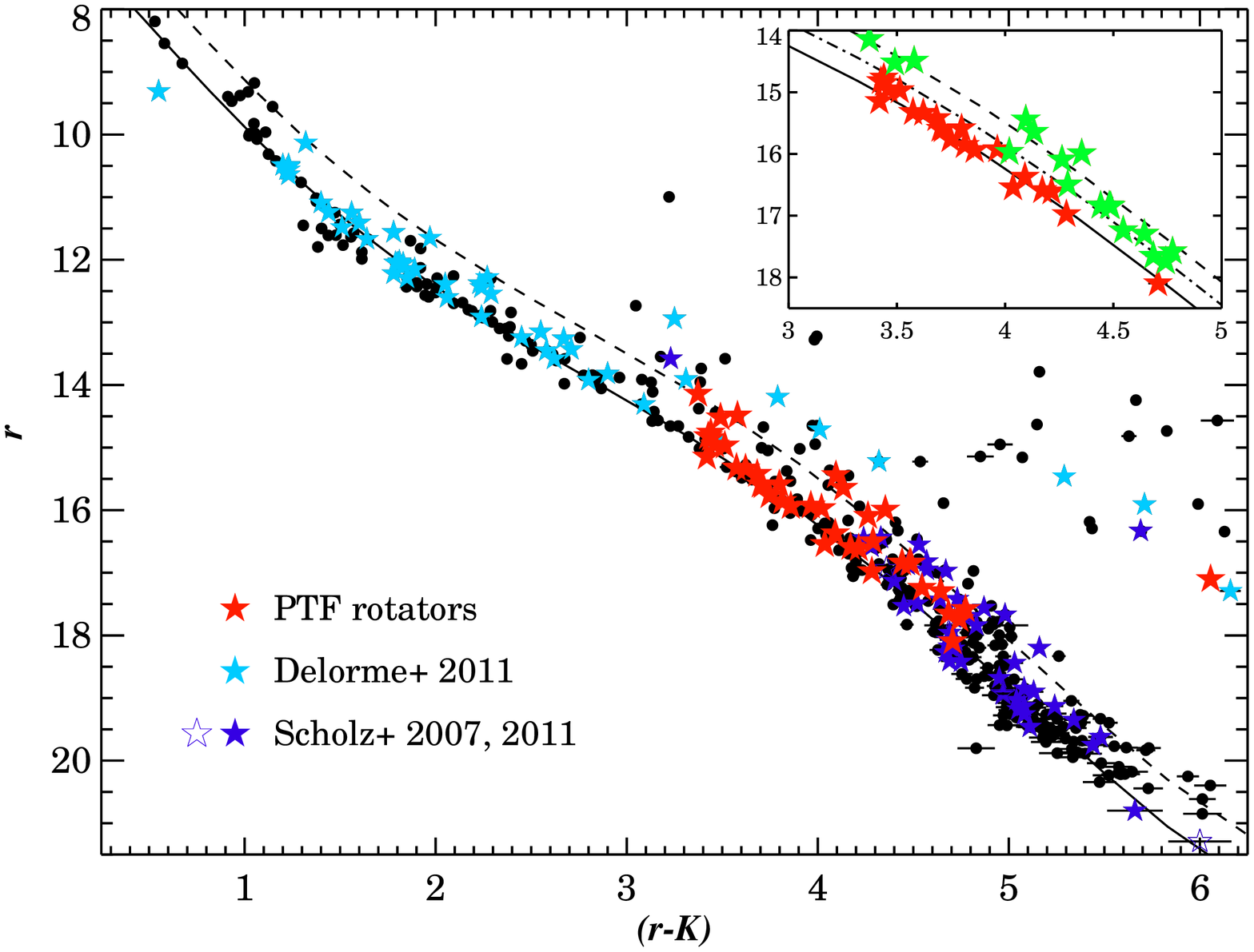}}
\caption{CMD with the Praesepe rotators identified by this study along with those of \citet{scholz2007}, \citet{scholz2011}, and \citet{delorme2011}. For cluster stars, we apply more stringent photometric tests, e.g., that SATURATED = 0 and that the $K$-band flag be ``A'' or ``B''; see \citet{stoughton02, 2mass}. To the single-star main sequence plotted in Figure~\ref{detections}, we add a binary main sequence offset by $0.75$ mag. The inset is a close-up of the region occupied by our rotators; stars above the dot-dashed line are candidate binaries (as is JS 497, which has an $(r-K) = 6$). Other cluster stars are omitted here for clarity.} 
\label{CMD_2} 
\end{figure}

\subsection{Comparison to other surveys of Praesepe}
\citet{delorme2011} surveyed two Praesepe fields with the eight cameras of SuperWASP as part of their search for transiting exoplanets. This survey produced 60-70 usable images per night for 60 nights spread over 130 nights, and these authors searched for rotation periods between 1.1 to 20 d for cluster members within 10 deg of the cluster center. As SuperWASP is optimized to detect transits around nearby, bright stars, the 52 stars for which this survey measured rotation periods (46 of which have $P_{mem} > 95\%$) are significantly brighter than those in our sample, and there is only one \citet{delorme2011} star, JS 545, for which we also measured $P_{rot}$.

\begin{deluxetable*}{lccccccccccc}[h!]
\tablewidth{0pt}
\tablecaption{High confidence rotators \label{hicp}}
\tablehead{
\colhead{}     & \colhead{}   & \colhead{}    & \colhead{}    & \colhead{}          & \colhead{}      & \colhead{$M_1$}     & \colhead{$M_2$} & \colhead {Ave.\ PTF}    & \colhead{\# of}    & \colhead{$P_{rot}$}   & \colhead{}\\
\colhead{Name} & \colhead{RA} & \colhead{Dec} & \colhead{SpT} & \colhead{$P_{mem}$} & \colhead{M$_K$} & \colhead{($\Msun$)} & \colhead{($\Msun$)} & \colhead{$R$ (mag)} & \colhead{Obs.} & \colhead{(d)} & \colhead{Power} 
}
\startdata
HSHJ 428 & 08 42 37.61 & $+$19 59 18.91 & M3.5 & 99.5 & 7.10 & {\bf 0.27} & 0.31 & 17.90$\pm$0.05 & 358 & 1.69 & 69.60 \\
JS 687\tablenotemark{a} & 08 35 59.44 & $+$20 04 40.54 & M3.7 & 99.8 & 6.71 & {\bf 0.33} & 0.37 & 17.55$\pm$0.03 & 172 & 1.76 & 42.33 \\
JS 718\tablenotemark{a} & 08 40 04.16 & $+$19 24 50.27 & M3.4 & 99.7 & 6.68 & {\bf 0.33} & 0.37 & 17.37$\pm$0.03 & 448 & 2.62 & 113.45 \\
JS 667\tablenotemark{a} & 08 33 17.98 & $+$19 16 32.75 & M3.7 & 98.9 & 6.52 & {\bf 0.36} & 0.40 & 17.40$\pm$0.03 & 347 & 2.30 & 102.98 \\
HSHJ 87\tablenotemark{a} & 08 35 47.22 & $+$18 08 29.97 & M3.2 & 98.8 & 6.41 & {\bf 0.38} & 0.41 & 17.04$\pm$0.02 & 168 & 2.12 & 41.65 \\
2MASS 08505688$+$1936579 & 08 50 56.87 & $+$19 36 57.89 & M2.5 & 97.1 & 6.40 & {\bf 0.38} & 0.41 & 16.76$\pm$0.02 & 320 & 11.95 & 49.70\\
AD 3814\tablenotemark{a} & 08 50 49.84 & $+$19 48 36.51 & M3.4 & 97.9 & 6.36 & {\bf 0.39} & 0.42 & 17.01$\pm$0.02 & 363 & 7.43 & 65.47 \\
AD 2552 & 08 39 22.44 & $+$20 04 54.77 & M1.9 & 99.9 & 6.22 & {\bf 0.41} & 0.44 & 16.33$\pm$0.01 & 530 & 25.36 & 127.43 \\
JS 241\tablenotemark{a} & 08 38 37.47 & $+$19 15 28.67 & M3.2 & 99.8 & 6.11 & {\bf 0.43} & 0.46 & 16.66$\pm$0.02 & 526 & 0.52 & 55.32 \\
JS 365 & 08 40 30.58 & $+$19 55 58.80 & M2.7 & 99.8 & 6.10 & {\bf 0.43} & 0.46 & 16.42$\pm$0.01 & 537 & 17.04 & 153.22 \\
JS 644 & 08 47 09.10 & $+$18 11 37.30 & M2.7 & 97.3 & 6.13 & {\bf 0.43} & 0.46 & 16.37$\pm$0.01 & 180 & 8.32 & 40.63 \\
JS 95\tablenotemark{a} & 08 35 40.15 & $+$18 42 28.32 & M3.3 & 99.4 & 6.07 & {\bf 0.44} & 0.46 & 16.74$\pm$0.02 & 168 & 1.98 & 46.04 \\
JS 250\tablenotemark{a} & 08 38 51.03 & $+$19 51 02.03 & M3.0 & 99.5 & 6.07 & {\bf 0.44} & 0.46 & 16.60$\pm$0.02 & 534 & 35.85 & 75.85 \\
JS 525 & 08 43 12.92 & $+$18 31 50.83 & M1.9 & 99.2 & 5.99 & {\bf 0.45} & 0.48 & 16.18$\pm$0.01 & 178 & 22.60 & 38.89 \\
JS 506\tablenotemark{a} & 08 42 52.28 & $+$19 51 45.92 & M2.5 & 99.4 & 5.92 & {\bf 0.47} & 0.49 & 16.27$\pm$0.01 & 364 & 3.97 & 101.18 \\
JS 283 & 08 39 16.79 & $+$19 47 42.63 & M1.4 & 99.8 & 5.79 & {\bf 0.49} & 0.51 & 15.71$\pm$0.01 & 534 & 19.62 & 116.87 \\
JS 110 & 08 36 08.97 & $+$19 13 48.02 & M1.3 & 99.7 & 5.75 & {\bf 0.50} & 0.51 & 15.68$\pm$0.01 & 170 & 18.21 & 40.71 \\
JS 237 & 08 38 32.83 & $+$19 46 25.61 & M1.2 & 99.9 & 5.72 & {\bf 0.51} & 0.52 & 15.59$\pm$0.01 & 535 & 18.57 & 88.75 \\
JS 148\tablenotemark{a} & 08 36 48.95 & $+$19 18 59.30 & M1.8 & 99.8 & 5.66 & {\bf 0.52} & 0.53 & 15.78$\pm$0.01 & 532 & 1.17 & 189.58 \\
JS 415 & 08 41 10.52 & $+$18 16 07.05 & M1.7 & 99.3 & 5.67 & {\bf 0.52} & 0.52 & 15.75$\pm$0.01 & 356 & 9.20 & 89.30 \\
JS 457 & 08 41 50.05 & $+$19 39 34.68 & M1.4 & 99.7 & 5.60 & {\bf 0.53} & 0.53 & 15.43$\pm$0.01 & 532 & 18.90 & 140.76 \\
JS 468 & 08 41 59.35 & $+$19 44 45.11 & M1.2 & 99.8 & 5.60 & {\bf 0.53} & 0.53 & 15.42$\pm$0.01 & 530 & 16.77 & 73.70 \\
JS 174\tablenotemark{a} & 08 37 19.91 & $+$19 03 11.92 & M3.0 & 99.7 & 5.54 & {\bf 0.54} & 0.54 & 15.80$\pm$0.01 & 347 & 2.03 & 61.36 \\
JS 187 & 08 37 32.42 & $+$19 31 17.98 & M1.9 & 99.0 & 5.50 & 0.55 & {\bf 0.55} & 15.42$\pm$0.01 & 372 & 1.81 & 113.91 \\
JS 46 & 08 33 50.76 & $+$19 46 58.62 & M0.5 & 99.5 & 5.44 & 0.57 & {\bf 0.56} & 14.98$\pm$0.00 & 347 & 17.00 & 80.65 \\
JS 364 & 08 40 28.23 & $+$18 56 08.96 & K7.8 & 99.2 & 5.42 & 0.57 & {\bf 0.56} & 15.08$\pm$0.00 & 356 & 17.33 & 60.84 \\
JS 489 & 08 42 34.86 & $+$20 59 40.80 & M1.0 & 95.9 & 5.46 & 0.56 & {\bf 0.56} & 15.18$\pm$0.00 & 183 & 17.33 & 61.77 \\
JS 505 & 08 42 49.68 & $+$18 51 35.10 & M1.5 & 94.4 & 5.45 & 0.56 & {\bf 0.56} & 15.33$\pm$0.00 & 176 & 13.50 & 49.98 \\
JS 430\tablenotemark{a} & 08 41 24.46 & $+$20 07 49.52 & M2.5 & 99.6 & 5.35 & 0.58 & {\bf 0.57} & 15.78$\pm$0.01 & 538 & 1.40 & 95.15 \\
JS 634\tablenotemark{a,b} & 08 46 38.21 & $+$19 52 44.78 & M2.5 & 62.3 & 5.22 & 0.61 & {\bf 0.59} & 15.42$\pm$0.01 & 359 & 3.24 & 80.98 \\
JS 181 & 08 37 28.45 & $+$20 36 28.54 & M0.8 & 99.1 & 5.20 & 0.62 & {\bf 0.60} & 14.84$\pm$0.00 & 185 & 7.22 & 82.29 \\
JS 432 & 08 41 24.18 & $+$18 14 02.69 & M1.0 & 99.3 & 5.17 & 0.62 & {\bf 0.60} & 15.02$\pm$0.00 & 356 & 2.09 & 134.68 \\
JS 545 & 08 43 44.73 & $+$21 12 34.32 & M0.8 & 97.5 & 5.12 & 0.63 & {\bf 0.61} & 14.60$\pm$0.00 & 183 & 2.83 & 53.99 \\
2MASS 08492676$+$1831196 & 08 49 26.76 & $+$18 31 19.54 & M0.5 & 97.8 & 5.10 & 0.64 & {\bf 0.61} & 14.65$\pm$0.00 & 178 & 9.20 & 64.69 \\
KW 563\tablenotemark{a,c} & 08 39 09.86 & $+$19 46 58.93 & M2.0 & 99.8 & 5.06 & 0.65 & {\bf 0.62} & 15.30$\pm$0.01 & 539 & 4.33 & 132.19 \\
KW 570 & 08 39 41.03 & $+$19 59 28.81 & M0.4 & 99.9 & 5.03 & 0.65 & {\bf 0.62} & 14.59$\pm$0.00 & 465 & 4.18 & 189.34 \\
JS 497\tablenotemark{a} & 08 42 42.07 & $+$19 17 32.26 & K7.3 & 94.4 & 4.76 & 0.71 & {\bf 0.67} & 14.09$\pm$0.01 & 153 & 12.68 & 50.40 \\
HSHJ 15\tablenotemark{a} & 08 31 40.45 & $+$19 47 54.19 & M2.0 & 95.2 & 4.74 & 0.71 & {\bf 0.68} & 14.36$\pm$0.00 & 347 & 9.36 & 114.60 \\
JS 159\tablenotemark{a} & 08 36 56.25 & $+$18 57 47.97 & M0.9 & 99.6 & 4.62 & 0.74 & {\bf 0.70} & 14.36$\pm$0.00 & 275 & 4.81 & 95.55 \\
AD 1508\tablenotemark{a} & 08 31 29.87 & $+$20 24 37.49 & M0.1 & 98.3 & 4.48 & 0.77 & {\bf 0.74} & 14.07$\pm$0.00 & 142 & 1.55 & 63.48 
\enddata
\tablenotetext{a}{Candidate binary system.}
\tablenotetext{b}{$P_{mem} = 62.3\%$; all other stars have $P_{mem} > 94\%$ \citep{adam2007}.}
\tablenotetext{c}{Listed as the variable V$*$AXCnc in SIMBAD.}
\tablecomments{Spectral types and membership probabilities are taken from \citet{adam2007}; for the former, the formal uncertainty is 0.1 spectral classes. However, the systematic uncertainty in the underlying definition of spectral types is $\sim$0.5 spectral classes for M dwarfs, and this systematic uncertainty will be reflected in the color-spectral type relations used for SED fits. M$_K$ is calculated assuming a distance of 181.5 pc \citep{vanleeuwen2009}. $M_1$ is estimated using the empirical \citet{delfosse2000} relation, while $M_2$ is estimated using the theoretical model of \citet{dotter2008}. Although the \citet{delfosse2000} relation extends to stars with M$_K$ = 4.5 mag, the predicted mass values diverge by up to about $5\%$ from those of \citet{dotter2008} for stars brighter than M$_K$ = 5.5 mag. The number in bold is the mass we assigned for a given star. We provide the mean magnitude of each light-curve after filtering on flags has been applied; the quoted error is the mean of the $1\sigma$ errors on the magnitudes used in calculating this mean magnitude.}
\end{deluxetable*}

The overlap between our sample and those of \citet{scholz2007} and \citet{scholz2011} is more significant. \citet{scholz2007} published the first rotation periods for Praesepe, for five mid- to late-M-type cluster members. These data were based on 125 images of a 0.36 deg$^2$ field taken over the course of about a month in early 2001 with the 2-m Schmidt telescope at the Th{\"u}ringer Landessternwarte Tautenburg (Germany), and on 108 images obtained over the course of a week two years later with the 3.5-m telescope at Calar Alto Observatory (Spain) of a different 0.25 deg$^2$ Praesepe field. One star for which they measured $P_{rot}$, JS 687, appears in our sample of rotators.

\citet{scholz2011} observed four $0.32$ deg$^2$ Praesepe fields around the cluster center with the 2.5-m Isaac Newton Telescope (Spain), obtaining a few hundred images for each field over the course of eight nights. This survey resulted in rotation periods for 49 stars, all but one of which have cluster $P_{mem} \geq\ 95\%$. These stars are typically fainter than those in our sample, but there are seven for which we measured $P_{rot}$.

\begin{deluxetable}{lccccc}[h!]
\tablewidth{0pt}
\tabletypesize{\scriptsize}
\tablecaption{Stars with previously reported periods \label{overlap}}
\tablehead{
\colhead{}     & \colhead{} & \multicolumn{2}{c}{$M$ ($M\sun$)}  & \multicolumn{2}{c}{$P_{rot}$ (d)} \\       
\colhead{Name} & \colhead{SpT}     & \colhead{Lit.} & \colhead{PTF}  & \colhead{Lit.} & \colhead{PTF} 
}
\startdata
HSHJ 428	& M3.5	& 0.26	& 0.27	& 1.71 & 1.69 \\
JS 687\tablenotemark{a}	& M3.7	& 0.29	& 0.33	& 1.76 & 1.76 \\
JS 718\tablenotemark{a}	& M3.4	& 0.33	& 0.33	& 2.65 & 2.62 \\
JS 241\tablenotemark{a}	& M3.2	& 0.42	& 0.43	& 0.53 & 0.52 \\
JS 506\tablenotemark{a}	& M2.5	& 0.47	& 0.47	& 3.97 & 3.97 \\
JS 430\tablenotemark{a}	& M2.5	& 0.54	& 0.57	& 1.37 & 1.40 \\
JS 545	& M0.8	& \nodata & 0.61 & 2.83 (0.004)\tablenotemark{b} & 2.83\\
KW 563\tablenotemark{a}	& M2.0	& 0.60	& 0.62	& 4.85 & 4.33 \\
KW 570	& M0.4	& 0.60	& 0.62	& 4.27 & 4.18 
\enddata
\tablenotetext{a}{Candidate binary system.}
\tablenotetext{b}{$P_{rot}$ is the average of the two periods whose difference is indicated in parentheses.}
\tablecomments{Data for JS 687 are from \citet{scholz2007} and for JS 545 from \citet{delorme2011}; all other data are from \citet{scholz2011}.}
\end{deluxetable}

The nine stars with published $P_{rot}$ values for which we have measured a rotation period are listed in Table~\ref{overlap}. The agreement between our values for these stars' periods and those in the literature is excellent, with differences $\gapprox$$2\%$ only for the two stars with $P_{rot} > 4$~d, KW 563 and KW 570. These are among the stars with the longest $P_{rot}$ reported by \citet{scholz2011}; neither is among the 24 stars flagged as having the most robust $P_{rot}$ measured by these authors, likely because of the difficulty of accurately measuring periods longer than half the duration of a monitoring campaign.

Our data occupy a unique area in mass-period space. The stars for which we measure $P_{rot}$ are later than those identified as rotators by \citet{delorme2011} and generally earlier than those identified by \citet{scholz2007} and \citet{scholz2011}, filling the gap between solar-type and mid-to-late M dwarf rotators in Praesepe. Furthermore, the cadence and time-span of our observations gives us sensitivity to the fast rotators identified by both of these groups and to the slow rotators identified at the high-mass end by \citet{delorme2011}. 

\subsection{Identifying potential binary systems}\label{bin}
The presence of binary systems could affect our interpretation of the Praesepe mass-period relation in two ways. First, treating a binary system as though it is a single star will lead to an erroneous mass estimate. Second, close binaries might be tidally locked and rotating faster than single stars of the same spectral type.\footnote{Some spot configurations could also lead to errors in our $P_{rot}$ measurements. As pointed out by \citet{scholz2011}, two spots of the same size separated by 180 deg in longitude would lead to a measurement of half the true stellar rotation period, for example.}

We use the cluster CMD to identify candidate binary systems among our rotators. \citet{steele1995} showed that in the Pleiades the effect of binaries on the CMD is to create a second ``main sequence'' lying above that of single stars and offset by $\sim$0.75 mag for a given color; we apply this same offset to the main sequence we derive from our fit to the \citet{adam2007} stellar SEDs (see Figure~\ref{CMD_2}). Stars along this sequence are likely to be in close to equal mass systems, with systems with more extreme mass ratios lying in the intervening region \citep{steele1995, hodgkin1999}. 

We adopt the method of \citet{hodgkin1999} and identify 18 candidate binary systems lying above the midpoint between the single-star and binary main sequences. These are highlighted in Figure~\ref{CMD_2} and in Table~\ref{hicp}. We note that the corresponding binary fraction is consistent (if slightly below) what has been reported previously for the cluster \citep[e.g.,][]{scholz2011}. Radial velocity measurements are required to confirm that these are binaries.

\begin{figure}
\centerline{\includegraphics[angle=-270,width=\columnwidth]{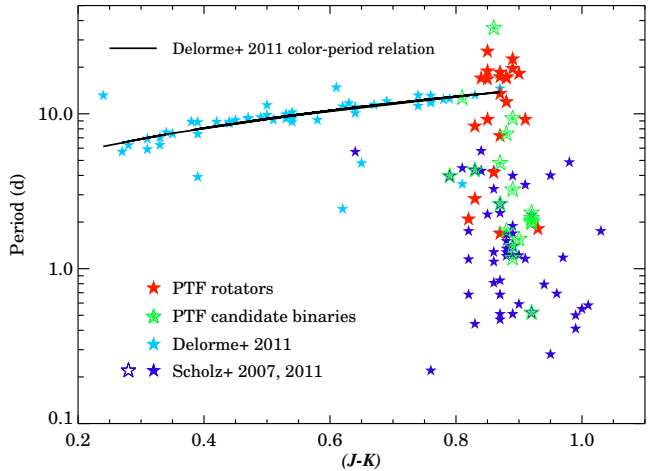}}
\caption{Color-period distribution for Praesepe, with the \citet{delorme2011} color-period relation overplotted.}
\label{cp}
\end{figure}

\begin{figure}[t!]
\centerline{\includegraphics[angle=-270,width=\columnwidth]{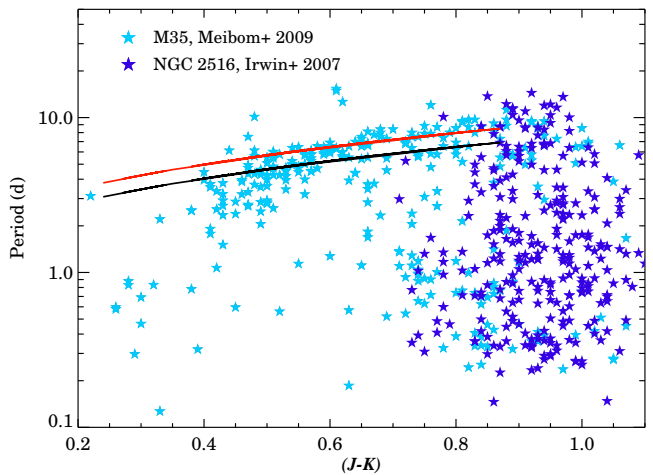}}
\caption{Color-period distribution for M35 and NGC 2516. The \citet{delorme2011} color-period relation shown in Figure~\ref{cp} has been spun up by $(600/150)^{0.5}$, following \citet{skumanich72}, to produce the black color-period relation. The red line takes the same \citet{delorme2011} relation and spins it up by $(600/150)^{0.35}$.}
\label{young}
\end{figure}

\section{Discussion}\label{results}
\subsection{Color-period relation for solar-type stars \& comparison to younger clusters}\label{cpr}
In Figure~\ref{cp}, we show the color-period distribution for Praesepe rotators, adding our new PTF data to the data collected by \citet{scholz2007}, \citet{delorme2011}, and \citet{scholz2011}. These data confirm that the color-period relation for Praesepe can be considered single-valued for stars bluer than $(J-K) \sim 0.8$, as found by \citet{delorme2011}. For redder, lower-mass stars, however, the distribution of stars in color-period space is scattered, with populations of fast and slow rotators. In particular, the PTF data allow us to measure $P_{rot} > 10$ d in stars with 0.8 $\lapprox\ (J-K)\ \lapprox$ 0.95, so that the sharp break in the color-period distribution noted by \citet{scholz2011} is no longer evident.\footnote{Long-term monitoring of the faintest Praesepe members is needed to determine whether the absence of slow rotators with $(J-K)\ \gapprox\ 1.0$ is real or an observational bias.} 

\citet{scholz2011} used the data collected by \citet{irwin2007} for the 150-Myr-cluster NGC 2516 to test models of rotational evolution by comparing the distribution of rotational periods in their Praesepe data to that for the younger cluster. We sought to replicate this analysis for both high- and low-mass stars; to this end we added to the \citet{irwin2007} data the rotation periods collected by \citet{meibom2009} for M35, another $\sim$150-Myr-cluster. For both clusters we required that the 2MASS photometry have either an ``A'' or ``B'' quality flag; this resulted in a sample of 300 stars for M35 and of 294 stars for NGC 2516. The resulting color-period plot for the two clusters is shown in Figure~\ref{young}.

We tested the \citet{skumanich72} law for the higher-mass stars in both clusters by spinning up the color-period relation derived by \citet{delorme2011} for Praesepe by $(600/150)^{0.5}$. The resulting color-period relation is plotted in Figure~\ref{young}. This spun-up color-period relation is fairly insensitive to the exact ages used for the clusters, and does not appear to describe the M35 stars well regardless of what (reasonable) age combination is used for the two clusters. Our analysis suggests that the the age dependence is closer to $t^{0.35}$ for the slow rotators with 0.5 $\lapprox\ (J-K)\ \lapprox$ 0.8; furthermore, it is difficult to find a single-valued color-period relation that describes these slow rotators and their M35 neighbors bluer than $(J-K) \sim 0.5$. 

While the spin-down derived from this comparison is less steep than that found by applying the \citet{skumanich72} law, \citet[][]{collier2009} find that $P_{rot} \propto t^{0.56}$ for (slightly earlier) solar-mass stars between the ages of the Hyades/Coma Ber (another $\sim$600-Myr-old cluster) and the age of the Sun. This implies that these stars are spinning down faster once they reach ages $>$600 Myr, and is consistent with e.g., the theoretical picture presented by \citet{irwin2009}: if the radiative core rotates more rapidly than the convective envelope when the star is very young, the transfer of angular momentum from core to envelope will ``soften'' the rotational braking of the surface layers early on. This will produce a sub-\citet{skumanich72} spin-down between $\sim$150 and 600 Myr, with the transition to a steeper spin-down rate occurring only once the transfer of angular momentum ends \citep[see Figure 5 of][]{irwin2009}.

For the color range we were most sensitive to with our PTF observations, however, working in color-period space was somewhat unsatisfying, as our sample spans a relatively narrow range of $(J-K)$ colors (as do those of \citet{scholz2007} and \citet{scholz2011}). We therefore chose to examine the relationship between masses and periods for these samples; this was further motivated by the fact that mass, unlike color, does not change significantly as the star evolves from the pre-main sequence onward.

\subsection{Masses, gyrochrones, \& mass-period comparisons}\label{mass}
A number of previous studies have used a star's $(B-V)$ as a proxy for its mass, and by calibrating the resulting color-period sequence in clusters of various ages, derived analytic expressions for a star's $P_{rot}$ as a function of $(B-V)$ and age \citep[e.g.,][]{barnes2003, mamajek2008}. Many of the lowest mass Praesepe members lack accurate $(B-V)$ colors, however. Furthermore, placing the Praesepe rotators and rotators from other clusters on a common mass scale requires a mass proxy with a large dynamic range. As shown in Figure~\ref{CMD_2}, $P_{rot}$ has been measured for Praesepe members spanning $\sim$13 mag in $r$, an achievement only made possible by combining light-curves obtained by facilities with distinctly different sensitivities. 

Fortunately, this span is somewhat compressed in the near infrared, as demonstrated by the range of $(r-K)$ colors covered by the Praesepe cluster sequence. To extend our analysis to the low-mass Praesepe cluster members, we therefore use mass estimates based on 2MASS photometry, which is available for all of the clusters we consider. Furthermore, as noted by \citet{delfosse2000}, near-infrared absolute magnitudes are better correlated with mass than their optical equivalents, at least for M dwarfs.

We began by calculating each star's absolute $K$ magnitude (M$_K$), using a distance to Praesepe of 181.5 pc \citep{vanleeuwen2009}. The empirical M$_K$-mass relation is the best calibrated of the absolute magnitude-mass relations described by \cite{delfosse2000}, and for stars with M$_K > $ 5.5, we derived masses using this relation. For stars with M$_K \leq$ 5.5, we estimated masses using the theoretical M$_K$-mass relation developed by \citet{dotter2008} for a 600 Myr, solar-metallicity population. This relation agrees well with the empirical M$_K$-mass relation of \citet{henry1993} but extends to higher masses. For completeness, both mass estimates are listed in Table~\ref{hicp} for each of our rotators; the adopted mass estimate is highlighted in boldface.

We followed the same steps to estimate the masses of Praesepe members with rotation periods measured by \citet{scholz2007}, \citet{delorme2011}, and \citet{scholz2011}.  Comparing our mass estimates to those derived by \citet{scholz2007} and \citet{scholz2011} from isochrone models indicates that the two techniques produce mass estimates that are consistent to within a few percent. Errors in the assumed cluster distance propagate linearly into the derived masses, so the uncertainties in our mass estimates reflect the $\sim$5\% uncertainty in the cluster distance \citep{vanleeuwen2009}; systematic uncertainties in the \citet{delfosse2000} relation are also of order $\sim$5$-$10\%, and we therefore adopt 10\% as the typical uncertainty in our derived masses. The location of the full sample in mass-period space is shown in Figure~\ref{mp}.

\subsubsection{Comparisons to model gyrochrones}
A major motivation for surveying stellar rotation in open clusters is to calibrate the relationship between a star's rotation period and its age. We have therefore used the formalism of \citet{barneskim2010} and \citet{barnes2010} to compute gyrochrones that quantify this relationship for a given age and zero age main sequence (ZAMS) rotation period. Eq.\ 22 in \citet{barnes2010} relates a star's convective turnover time ($\tau$) to a quadratic function of the star's age and the ratio between its ZAMS and present day rotation period. Adopting an age of 600 Myr for Praesepe, we follow \citet{barnes2010} in calculating gyrochrones for stars with ZAMS periods ranging from 3.4 days down to 0.12 days, corresponding to break-up for a solar-type star.

\begin{figure}
\centerline{\includegraphics[angle=-270,width=\columnwidth]{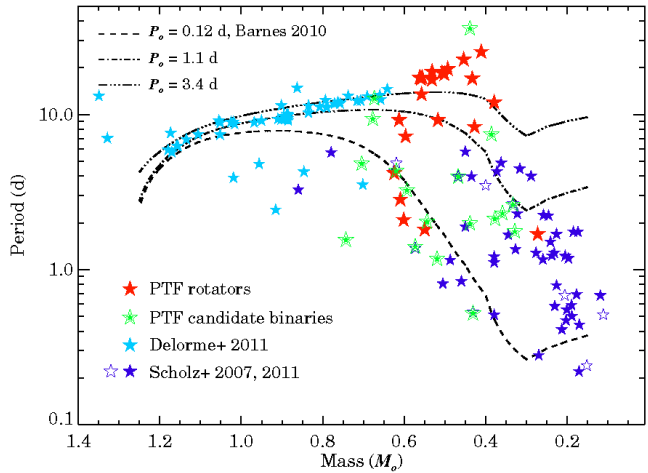}}
\caption{Mass-period distribution for Praesepe, with model gyrochrones from \citet{barnes2010} overplotted. Each gyrochrone corresponds to the predicted position at 600 Myr of stars of a range of masses but with the same ZAMS rotation period, $P_o$.}
\label{mp}
\end{figure}

\begin{figure}
\centerline{\includegraphics[angle=-270,width=\columnwidth]{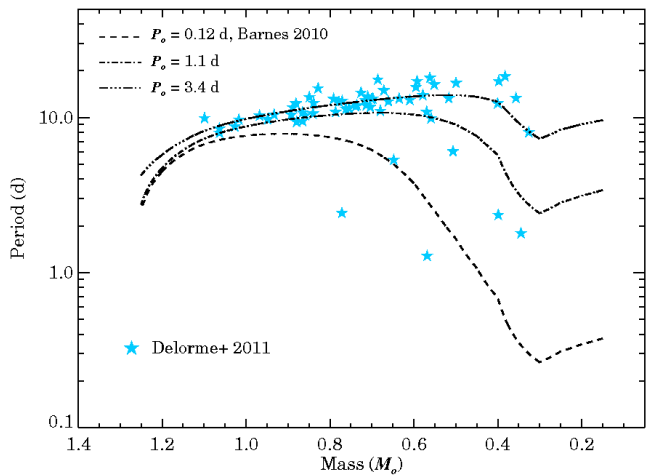}}
\caption{Mass-period distribution for the Hyades, again with gyrochrones for a 600 Myr population overplotted.}
\label{hyads}
\end{figure}

When age and ZAMS period are fixed, the \citet{barnes2010} quadratic equation becomes a double-valued relationship between $\tau$ and current-day period. Using the relation between stellar mass and global convective turnover timescale tabulated by \citet{barneskim2010} to project the $\tau$-period relationships into mass-period space, we overlay in Figure~\ref{mp} the resulting 600 Myr gyrochrones. These gyrochrones bound the region of the mass-period plane containing Praesepe rotators reasonably well, indicating a tight mass-period relation for masses $>$0.7 $\Msun$ and a significantly broader range of periods at masses $<$0.6 $\Msun$. The sample of Praesepe rotators, however, includes several stars that lie outside the region of the mass-period plane enclosed by these gyrochrones, suggesting either that Praesepe stars had a broader range of ZAMS $P_{rot}$ than 0.12-3.4 days, or that this analytic description does not yet fully replicate the morphology of the $\sim$600 Myr mass-period relation. 

\subsubsection{Comparison to the Hyades}
We have collected observational data for candidate members of the Hyades whose $P_{rot}$ were measured by \citet{delorme2011}. We estimated masses for these stars following the technique described above and using a distance to the Hyades of 46.45 pc \citep{vanleeuwen2009}. The Hyades is sufficiently close that many members have direct parallax measurements: for those, we adopted the distances implied by their individual parallaxes as reported by \citet{delorme2011}. The resulting mass-period plot is shown in Figure~\ref{hyads}.

Figures~\ref{mp} and \ref{hyads} suggest that the transition from a slow-rotating to mixed fast- and slow-rotating population occurs at a lower mass in the Hyades than in Praesepe. While the data for low-mass stars are sparser than for Praesepe, there are comparatively very few fast rotators in the Hyades above a mass of $\sim$0.5 $\Msun$. This strengthens the \citet{delorme2011} conclusion that the clusters differ in age by $\sim$50 Myr, with Praesepe being the younger of the two.

\subsection{The evolution of the mass-period relation}
These comparisons of a stellar population to gyrochrones generated for a given age are one test of semi-empirical models' ability to describe the angular momentum evolution of stars. Another is to compare the model predictions for a star of a given mass to the data available for stars of that mass at various ages. We selected Praesepe stars not flagged as potential binaries, separated them into two mass bins, 0.3 $\leq M <$ 0.5 $\Msun$ and 0.5 $\leq M <$ 0.7 $\Msun$, and, following \citet{scholz2011}, calculated the median and 10$^{\rm th}$ and 90$^{\rm th}$ percentile $P_{rot}$ for these two samples ($P_{rot}$ = 3.98, 0.84, and 19.62 d, respectively, for the low-mass bin; $P_{rot}$ = 13.09, 1.81, and 18.21 d for the high-mass one). We then used the \citet{barnes2010} models to find the corresponding ZAMS $P_o$ for each of these representative 600 Myr stars, to which we assign masses of $0.4$ and $0.6$ $\Msun$. This $P_o$ was fed back into the models to predict the $P_{rot}$ of these representative stars at ages ranging from 125 Myr to 10 Gyr, and including 150 Myr, 650 Myr, 1.5 Gyr, and 8.5 Gyr. 

\begin{figure}[h!]
\centerline{\includegraphics[angle=-270,width=\columnwidth]{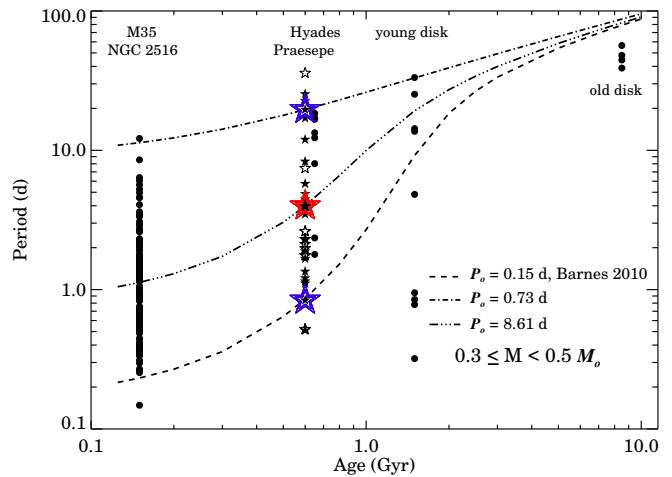}}
\caption{Predicted evolutionary tracks for 0.4 $\Msun$ stars in Praesepe compared to the observed spread in $P_{rot}$ in populations of different ages. The red star is the median  Praesepe rotator for 0.3 $\leq$ M $<$ 0.5 $\Msun$, while the blue stars are the median and 10$^{\rm th}$ and 90$^{\rm th}$ percentile rotators. The $P_o$ for these stars and evolutionary tracks are calculated using the models of \citet{barnes2010}. Candidate binaries flagged in \S~\ref{bin} are not included in these calculations but are plotted as empty symbols. }
\label{agep1}
\end{figure}

\begin{figure}[h!]
\centerline{\includegraphics[angle=-270,width=\columnwidth]{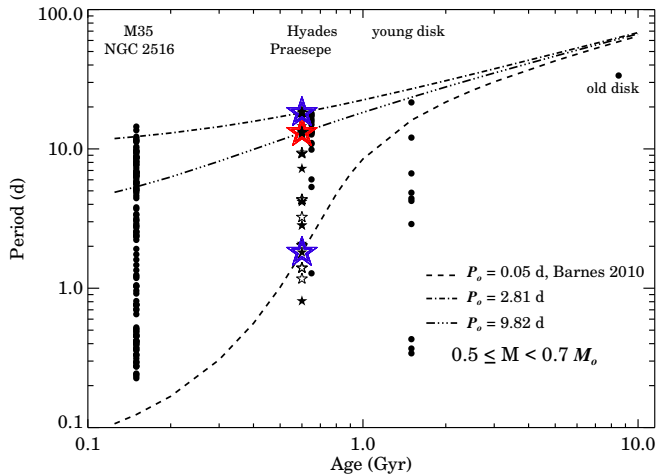}}
\caption{Same as in Figure~\ref{agep1}, for 0.6 $\Msun$ stars in Praesepe.}
\label{agep2}
\end{figure}

The resulting evolutionary tracks are plotted in Figures~\ref{agep1} and \ref{agep2}, along with period data for these mass bins from M35, NGC 2516, the Hyades, and young and old disk stars \citep[1.5 and 8.5 Gyr;][]{kiraga2007}. The mass estimates for the 150-Myr-old clusters were done in the manner described in \S~\ref{mass}.\footnote{\citet{irwin2007} provide mass estimates for NGC 2516 based on comparisons to the models of \citet{baraffe98}. These two sets of estimated masses agree to within a few hundredths of a solar mass, with our masses being typically slightly lower.} For the \citet{kiraga2007} stars, we used the mass estimates these authors provide, which are based on the empirical \citet{delfosse2000} relation for M$_V$.

The agreement between the models for a $0.4$ $\Msun$ star and the observed distribution of rotation periods is good between the ages of 150 and 600 Myr. The periods predicted for the Praesepe median and 10$^{\rm th}$ and 90$^{\rm th}$ percentile $P_{rot}$ stars at 150 Myr fall within the range of periods observed in M35 and NGC 2516 for stars between 0.3 and 0.5 $\Msun$, although they are close to the extreme periods observed in these clusters. 

The \citet{barnes2010} models predict a faster spin-down as stars reach an age $>$1 Gyr than what is seen: by $1.5$ Gyr, the models have converged significantly, while the spread in the observed $P_{rot}$ remains roughly equivalent to that seen at 600 Myr. Extending the models to the age of the thick disk, the rotation periods predicted for all percentiles are a few times the (few) periods measured for old disk stars. However, the \citet{kiraga2007} sample is selected from known X-ray-luminous stars, and this may bias this sample toward faster rotators. In addition, the young and old disk samples are not single-age populations; their age spreads are such that they likely include stars spanning the full range of ages shown here. If the fast rotators are systematically younger than the slow ones, the true disagreement with the models is not as significant as suggested by Figure~\ref{agep1}. 

The agreement between the models and the observed spread in $P_{rot}$ for the representative 0.6 $\Msun$ stars is also good between 150 and 600 Myr, with one clear exception. Replicating the 10$^{\rm th}$ percentile rotator in Praesepe requires a ZAMS 0.6 $\Msun$ star with $P_o = 0.05$ d, which implies that this star is rotating faster than its breakup $P \approx 0.06$ d \citep[assuming that a 0.6 $\Msun$ star has R $= 0.56\ \Rsun$;][]{dotter2008}. At later ages, the models converge even more than for the less massive star, so that the match between predicted period distribution and that seen at ages $>$1 Gyr is worse. However, here again, the limitations of the disk star sample make comparisons to the models difficult to interpret.

\section{Conclusion}\label{concl}
Using PTF observations of four overlapping fields, we have measured rotation periods for 40 late-K to mid-M stars belonging to the nearby, rich, $\sim$600 Myr open cluster Praesepe. Our data occupy a unique area in mass-period space: we measure $P_{rot}$ for stars later than those identified as rotators by \citet{delorme2011} and generally earlier than those identified by \citet{scholz2007} and \citet{scholz2011}, filling the gap between solar-type and mid-to-late M dwarf rotators in Praesepe. Furthermore, the cadence and time-span of our observations gives us sensitivity to the fast rotators identified by these groups and to the slow rotators identified at the high-mass end by \citet{delorme2011}.

Our measurements indicate that Praesepe's mass-period relation undergoes a transition from a well-defined singular relation to a more scattered distribution of both fast and slow-rotators at masses $\sim$0.6 $\Msun$, corresponding roughly to a spectral type of M1. The location of this transition is broadly consistent with expectations based on observations of younger clusters and the assumption that stellar-spin down is the dominant mechanism influencing angular momentum evolution at $\sim$600 Myr. 

A comparison to the data recently published by \citet{delorme2011} for the Hyades, widely assumed to be coeval to Praesepe, suggests that this transition occurs at different characteristic masses in the two clusters, providing further evidence that Praesepe is the younger of the two clusters. Furthermore, by using the \citet{barneskim2010} and \citet{barnes2010} formalisms to evolve the Praesepe $P_{rot}$ in time and comparing the predicted $P_{rot}$ with measured $P_{rot}$ in M35 and NGC 2516 ($\sim$150 Myr) and for young and old field star populations (1.5 and 8.5 Gyr), we find that stellar spin-down may progress more slowly than described by these relations. 

\begin{deluxetable*}{lccccl}[t!]
\tablewidth{0pt}
\tablecaption{Other Interesting Variable Stars}
\tablehead{
\colhead{}       & \colhead {Ave.\ PTF}    & \colhead{\# of}    & \colhead{$P_{rot}$}   & \colhead{}   & \colhead{}\\
\colhead{SDSS J} & \colhead{$R$ (mag)} & \colhead{Obs.} & \colhead{(d)} & \colhead{Power}  & \colhead{Type}
}
\startdata
083113.96$+$194951.5 & $15.99\pm0.01$ & 327 & 0.26 & 136.59 & Eclipsing binary? \\
083238.89$+$210424.6\tablenotemark{a} & $17.44\pm0.03$ & 173 & 0.55 & 57.20 & RR Lyrae \\
083426.37$+$202040.9 & $15.53\pm0.01$ & 348 & 0.13 & 60.69 & W Uma? \\
083525.00$+$194659.7 & $15.96\pm0.01$ & 170 & 0.80 & 63.81 & RR Lyrae \\
083706.80$+$185556.2 & $16.14\pm0.01$ & 348 & 0.30 & 145.78 & W Uma? \\
083816.78$+$182724.6\tablenotemark{a} & $17.36\pm0.03$ & 348 & 0.29 & 108.32 & RR Lyrae \\
084041.70$+$201612.1 & $16.33\pm0.02$ & 377 & 0.13 & 123.29 & W Uma? \\
084214.42$+$195819.7 & $16.33\pm0.01$ & 531 & 0.25 & 197.15 & Eclipsing binary? \\
084803.57$+$191340.5 & $16.21\pm0.01$ & 180 & 0.49 & 55.37 & Known RR Lyrae\tablenotemark{b}\\
084920.49$+$182617.3 & $15.85\pm0.01$ & 181 & 0.53 & 63.28 & Known RR Lyrae\tablenotemark{a,b}\\
084950.50$+$193232.0 & $14.70\pm0.00$ & 337 & 0.18 & 137.16 & W Uma? \\
085112.62$+$184344.3 & $15.06\pm0.00$ & 181 & 0.19 & 84.19 & W Uma?
\enddata
\tablenotetext{a}{Spectrum is available from SDSS.}
\tablenotetext{b}{Classified as RR Lyrae in SIMBAD.}
\label{ap_table}
\end{deluxetable*}

The fixed age mass-period relation is but one projection of the underlying stellar age-rotation-activity relationship. Previous studies of stellar activity, in clusters and the field, have derived relationships between a star's age and observational tracers of its coronal or chromospheric activity \citep[e.g.,][]{skumanich72, radick1987, soderblom2001}. Recent studies of chromospheric activity in low-mass field stars have inferred activity lifetimes as a function of spectral type by modeling the vertical gradient in H$\alpha$ emission strengths as a consequence of dynamical heating in the Galactic disk \citep[e.g.,][]{andy08}. These studies predict that stars with spectral types of M2 or later have activity lifetimes $>$1 Gyr. The activity lifetimes of M0-M1 stars are somewhat less well known, as few active early M stars are observed in the field, but appear to be $\lesssim$600 Myr.

These relations would thus predict that the boundary between H$\alpha$ active and inactive Praesepe members should occur in the M0/M1 spectral range. The agreement between the implied mass of the active/inactive boundary in Praesepe, near 0.6 $\Msun$, and the similar characteristic mass for the transition from a singular mass-period relation to a more scattered distribution of rapid and slowly rotators, strengthens the case for an underlying rotation-activity relation in this cluster. In a forthcoming paper, we use the results of our spectroscopic campaign with the 2.4-m Hiltner telescope at MDM Observatory and the WIYN 3.5-m telescope at NOAO, both on Kitt Peak, AZ, to examine this relationship between rotation and activity in Praesepe.

\acknowledgments We thank Sydney Barnes for detailed discussions of his models, and Mark Giampapa, Joel Hartman, Jonathan Irwin, Steve Saar, and Aleks Scholz for their useful comments on a draft of the paper. We are grateful to Janet Jacobsen for her contributions to the PTF infrastructure; without her efforts, this work would not have been possible. Finally, we thank the referee for a very prompt review of this manuscript, and for several helpful suggestions that improved it.

K.R.C.\ and A.L.K.\ acknowledge support provided by NASA through Hubble Fellowship grants HST-HF-51253.01 and 51257.01 awarded by the STScI, which is operated by the AURA, Inc., for NASA, under contract NAS 5-26555.

This research has made use of NASA's Astrophysics Data System Bibliographic Services, the SIMBAD database, operated at CDS, Strasbourg, France, the NASA/IPAC Extragalactic Database, operated by the Jet Propulsion Laboratory, California Institute of Technology, under contract with the National Aeronautics and Space Administration, and the VizieR database of astronomical catalogs \citep{Ochsenbein2000}. IRAF (Image Reduction and Analysis Facility) is distributed by the National Optical Astronomy Observatories, which are operated by the Association of Universities for Research in Astronomy, Inc., under cooperative agreement with the National Science Foundation. 
 
The Two Micron All Sky Survey was a joint project of the University of Massachusetts and the Infrared Processing and Analysis Center (California Institute of Technology). The University of Massachusetts was responsible for the overall management of the project, the observing facilities and the data acquisition. The Infrared Processing and Analysis Center was responsible for data processing, data distribution and data archiving.  

\renewcommand{\thesection}{A\arabic{section}}
\setcounter{section}{0}  
\section*{Appendix: Interesting Variable Stars in the Praesepe Fields}
Our light-curve analysis was largely restricted to stars identified previously as candidate Praesepe members by \citet{adam2007}. Candidate Praesepe members total less than 1\% of the objects in our target fields, however, so that many other variable stars are likely to be present in the full catalog of PTF light-curves.

To explore the population of variable stars that can be detected by PTF at moderate Galactic latitude, we performed a search for high-confidence variables within this full catalog of light-curves. Candidate variables were identified by computing the ratios of the $\sigma$ of the raw light-curve to the $\sigma$ of light-curves that were boxcar smoothed over windows spanning 9 and 36 epochs. Boxcar smoothing light-curves dominated by random noise will reduce random fluctuations and shrink the $\sigma$ by $\sqrt{n}$. By contrast, light-curves dominated by structured variability will not converge with $\sqrt{n}$, so that the ratio of the $\sigma$s of smoothed and raw light-curves serves as a simple tool for identifying structured variability. As the light-curves presented here are irregularly sampled, the boxcar windows corresponded to different timescales at different points in the light-curve, and were sensitive to variability over a larger range of timescales than if the light-curves were regularly sampled.

By computing $\sigma$ ratios for stars with $R <18$ mag, we identified $\sim$6000 stars as candidate variables. We produced periodograms in the same manner as described in \S~\ref{periods} and defined stars with periodogram peaks $>$50 as are likely periodic variables. Visual inspection then identified 12 robust detections of large amplitude ($>$0.1 mag) variable stars. Phased light-curves for these stars are presented in Figure~\ref{ap_lcs} and the stars are tabulated in Table~\ref{ap_table}. These are nearly all short period ($P<1$ d) variables; classifying these on the basis of their light-curve shapes and amplitudes, we identify five as likely RR Lyrae stars (two of these have previously been identified as RR Lyrae) and two as candidate eclipsing systems. The remaining five show large amplitude sinusoidal variations, and on the basis of their short periods, we tentatively classify these objects as W Uma systems, but further study is necessary to confirm their status.

\begin{figure*}[th]
\centerline{\includegraphics[width=2\columnwidth]{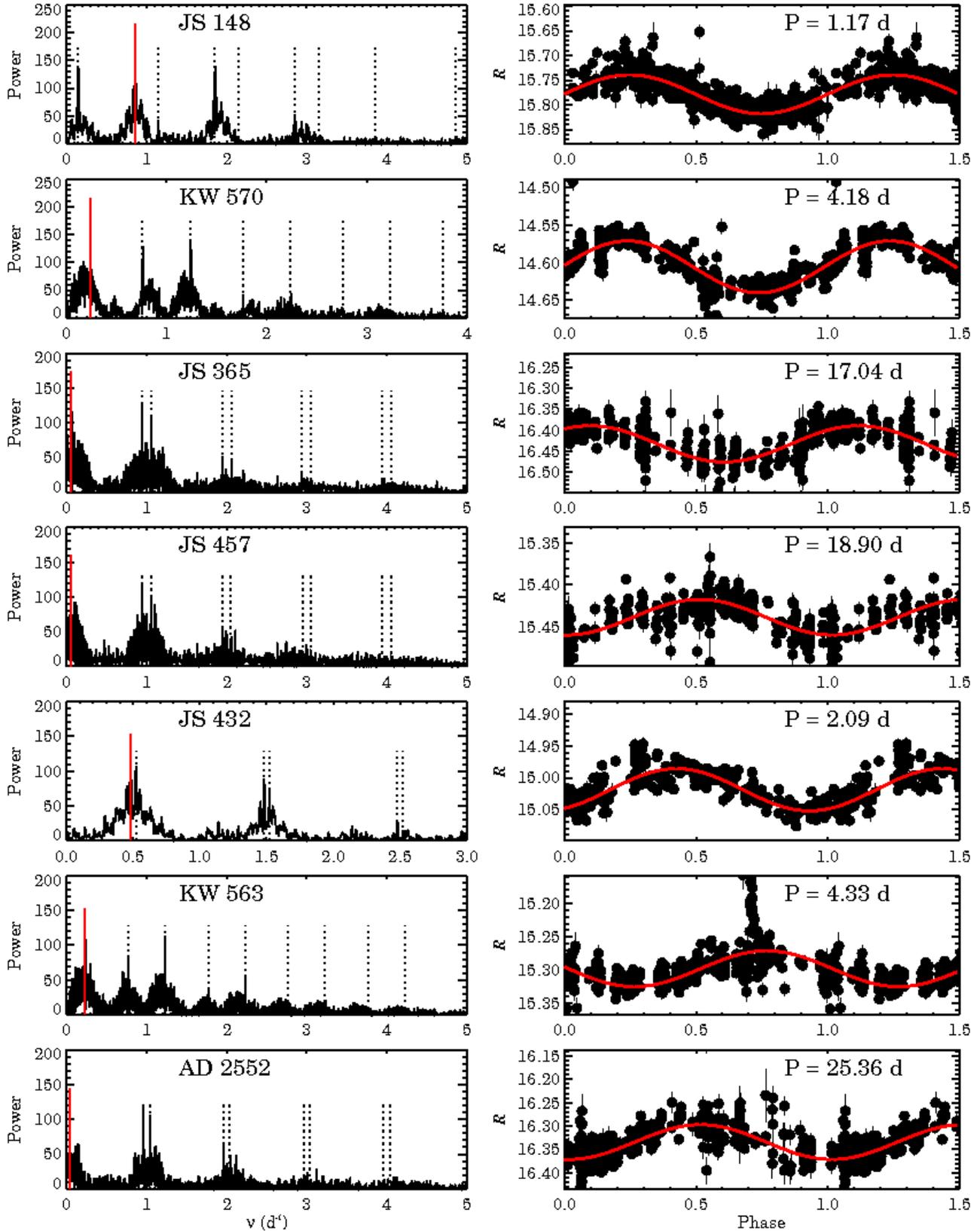}}
\caption{Periodograms (left column) and phased light-curves (right column) for each Praesepe member with a rotation period measured from our PTF data. The adopted $P_{rot}$ is flagged in each periodogram with a red line; potential beat frequencies between this period and an assumed one-day alias are flagged with black dotted lines. Error bars are overplotted on each data point in the light-curve, which is phased by $P_{rot}$. For clarity, we show 1.5 phases of each star's period and plot a sinusoid curve with the same amplitude and period as the variability measured for each star.}\label{image}
\end{figure*}

\begin{figure*}[th]
\centerline{\includegraphics[width=2\columnwidth]{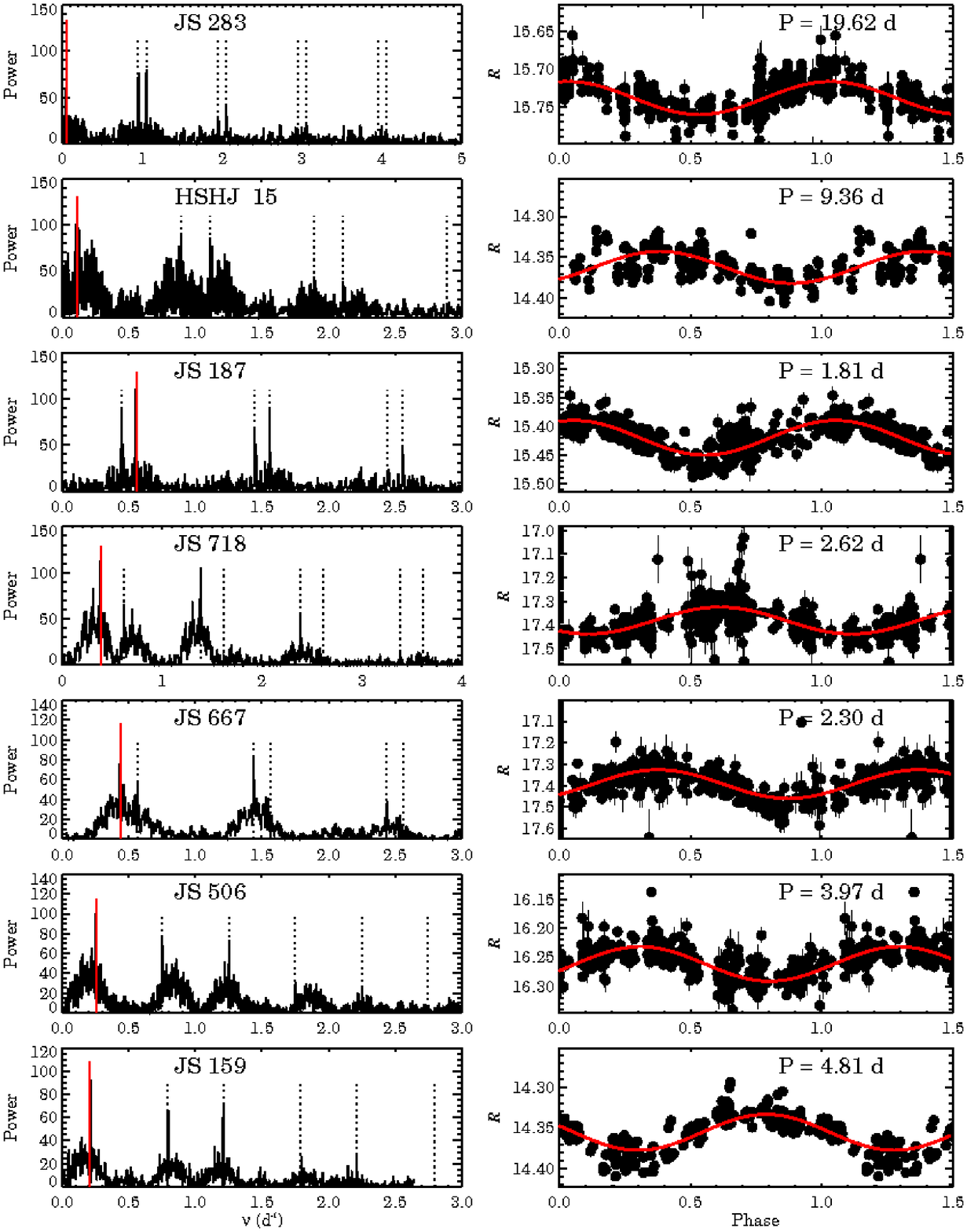}}
\end{figure*}

\begin{figure*}[th]
\centerline{\includegraphics[width=2\columnwidth]{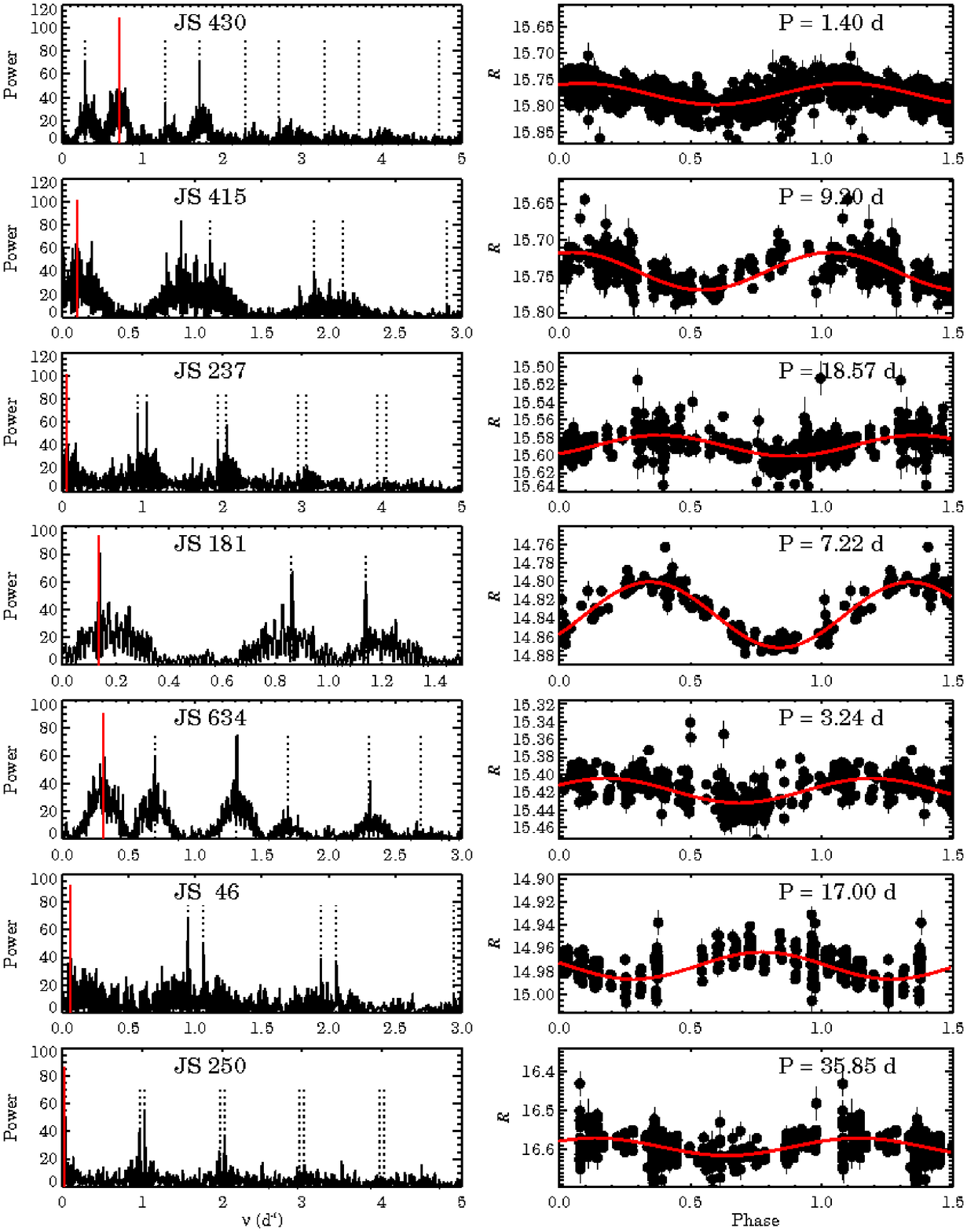}}
\end{figure*}

\begin{figure*}[th]
\centerline{\includegraphics[width=2\columnwidth]{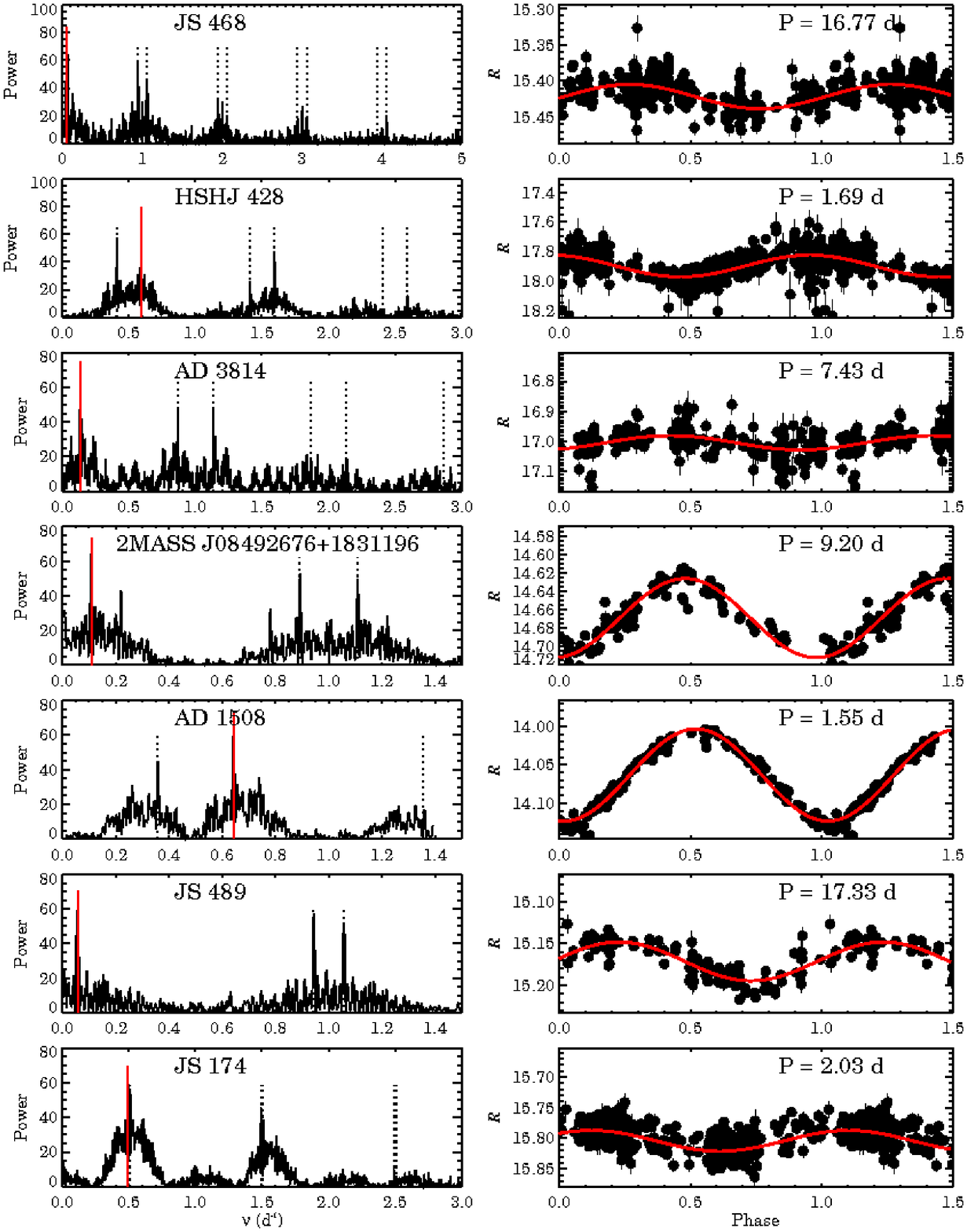}}
\end{figure*}

\begin{figure*}[th]
\centerline{\includegraphics[width=2\columnwidth]{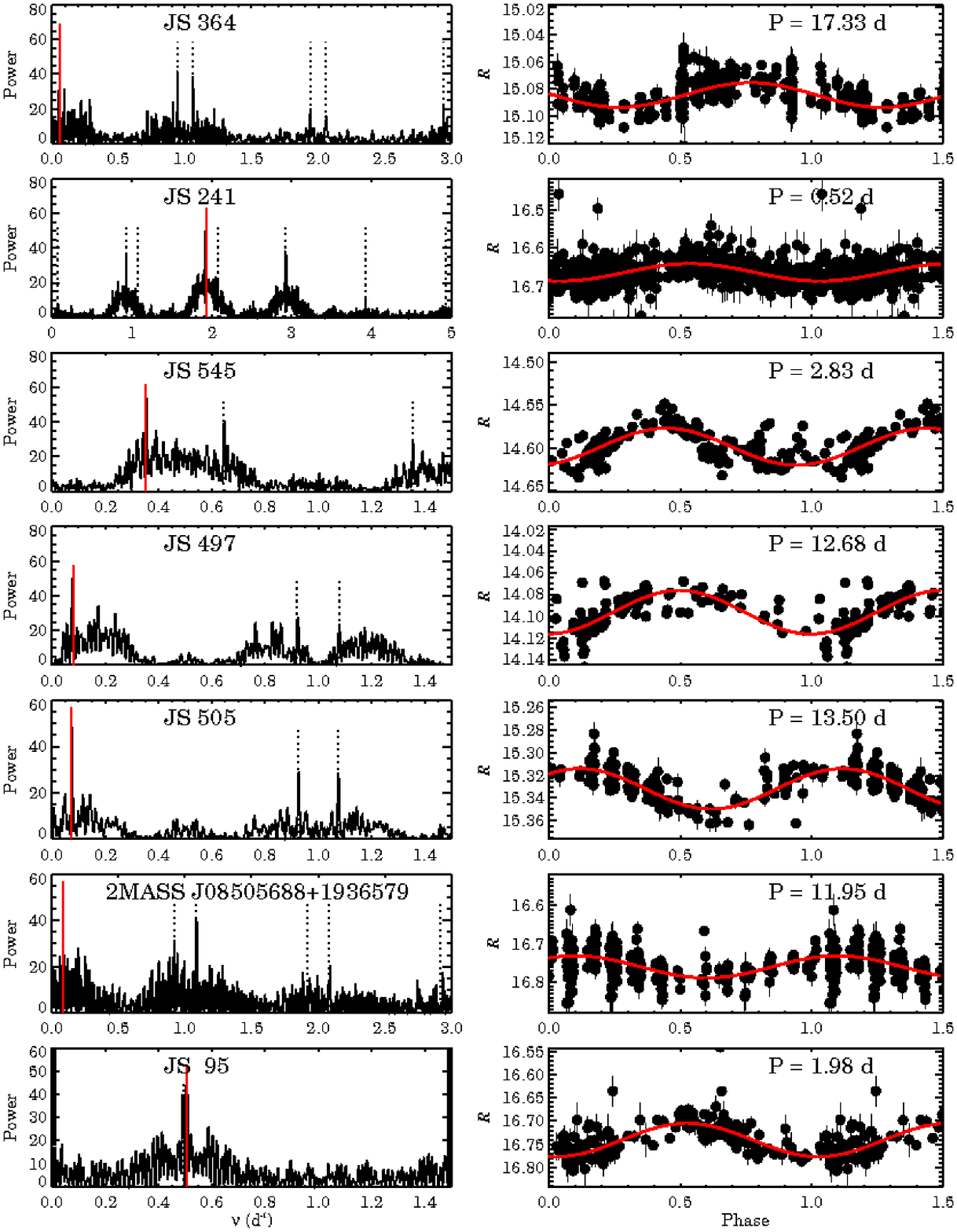}}
\end{figure*}

\begin{figure*}[th]
\centerline{\includegraphics[width=2\columnwidth]{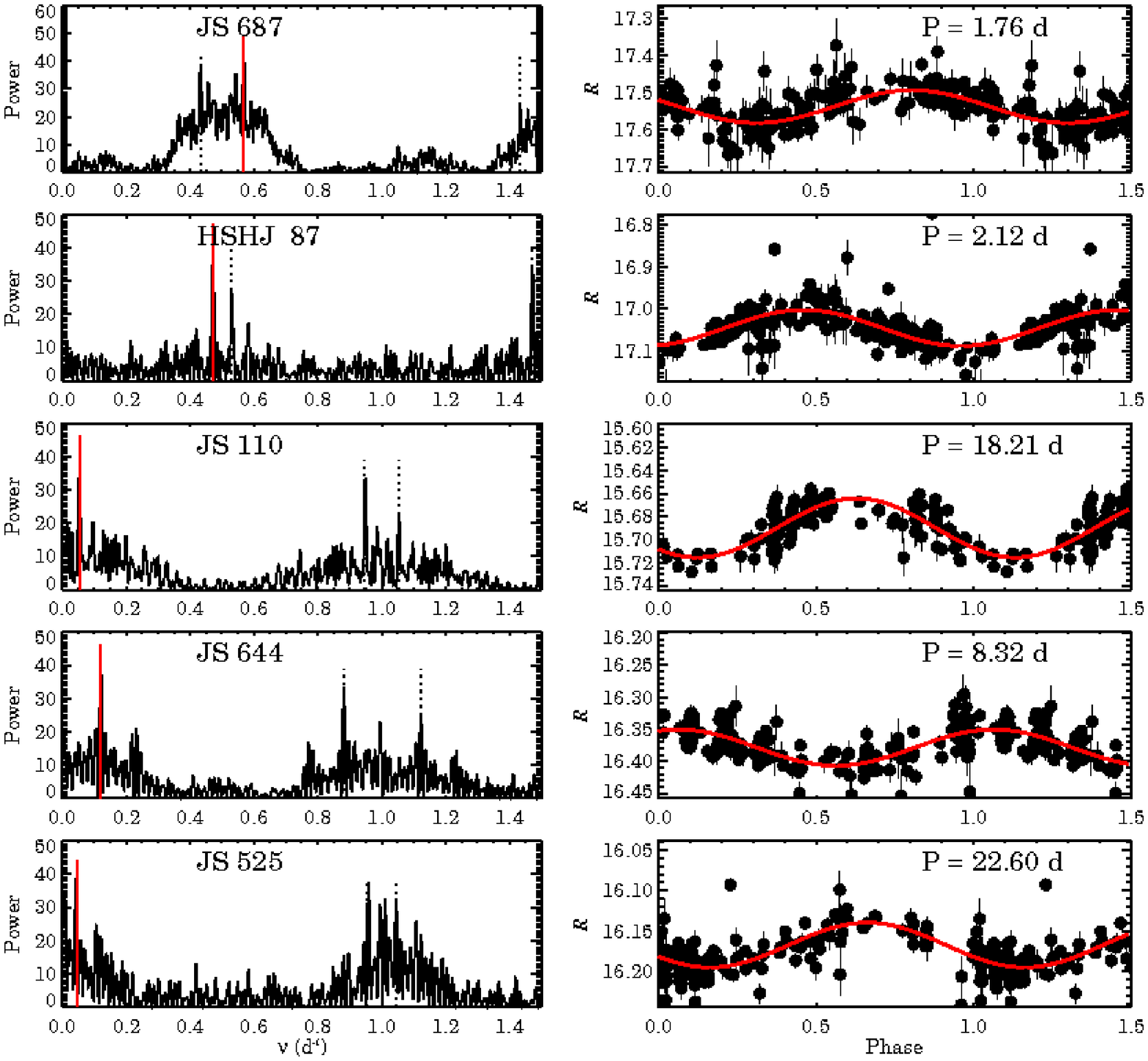}}
\end{figure*}

\begin{figure*}[th]
\centerline{\includegraphics[width=2\columnwidth]{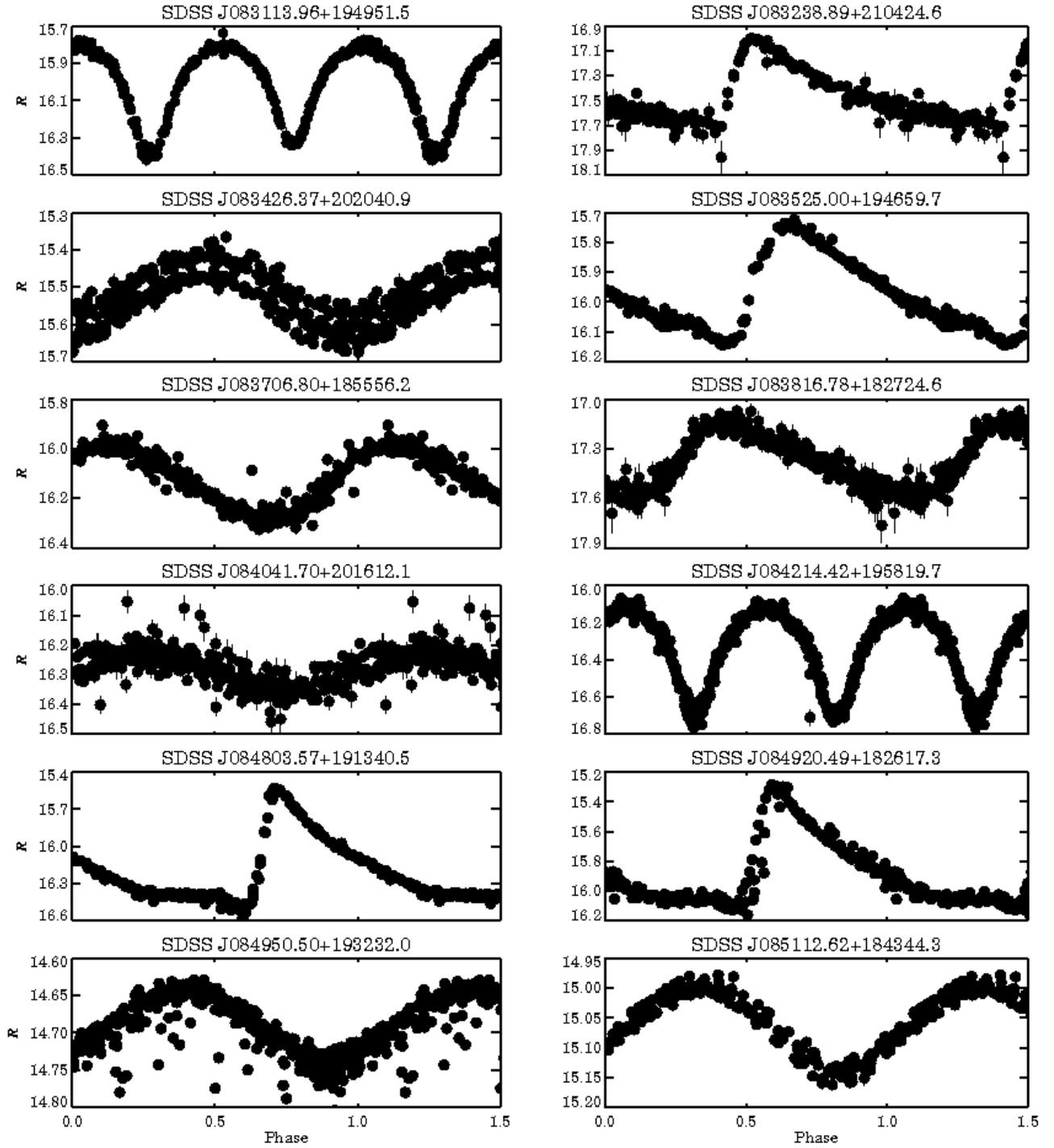}}
\caption{Other interesting periodic variables in the Praesepe fields. SDSS J084803.57$+$191340.5 and 084920.49$+$182617.3 are both previously identified RR Lyrae stars.}\label{ap_lcs}
\end{figure*}


\clearpage
\begin{thebibliography}{53}
\expandafter\ifx\csname natexlab\endcsname\relax\def\natexlab#1{#1}\fi

\bibitem[{{Baraffe} {et~al.}(1998){Baraffe}, {Chabrier}, {Allard}, \&
  {Hauschildt}}]{baraffe98}
{Baraffe}, I., {Chabrier}, G., {Allard}, F., \& {Hauschildt}, P.~H. 1998, \aap,
  337, 403

\bibitem[{{Barnes}(2003)}]{barnes2003}
{Barnes}, S.~A. 2003, \apj, 586, 464

\bibitem[{{Barnes}(2007)}]{barnes2007}
---. 2007, \apj, 669, 1167

\bibitem[{{Barnes}(2010)}]{barnes2010}
---. 2010, \apj, 722, 222

\bibitem[{{Barnes} \& {Kim}(2010)}]{barneskim2010}
{Barnes}, S.~A., \& {Kim}, Y.-C. 2010, \apj, 721, 675

\bibitem[{{Bertin} \& {Arnouts}(1996)}]{bertin96}
{Bertin}, E., \& {Arnouts}, S. 1996, \aaps, 117, 393

\bibitem[{{Browning}(2008)}]{Browning2008}
{Browning}, M.~K. 2008, \apj, 676, 1262

\bibitem[{{Collier Cameron} {et~al.}(2009){Collier Cameron}, {Davidson},
  {Hebb}, {Skinner}, {Anderson}, {Christian}, {Clarkson}, {Enoch}, {Irwin},
  {Joshi}, {Haswell}, {Hellier}, {Horne}, {Kane}, {Lister}, {Maxted}, {Norton},
  {Parley}, {Pollacco}, {Ryans}, {Scholz}, {Skillen}, {Smalley}, {Street},
  {West}, {Wilson}, \& {Wheatley}}]{collier2009}
{Collier Cameron}, A. {et~al.} 2009, \mnras, 400, 451

\bibitem[{{Covey} {et~al.}(2008){Covey}, {Ag{\"u}eros}, {Green}, {Haggard},
  {Barkhouse}, {Drake}, {Evans}, {Kashyap}, {Kim}, {Mossman}, {Pease}, \&
  {Silverman}}]{Covey2008}
{Covey}, K.~R. {et~al.} 2008, \apjs, 178, 339

\bibitem[{{Covey} {et~al.}(2010){Covey}, {Agueros}, {Lemonias}, {Law}, \&
  {Kraus}}]{covey2010}
{Covey}, K.~R., {Agueros}, M.~A., {Lemonias}, J.~J., {Law}, N.~M., \& {Kraus},
  A.~L. 2010, ArXiv e-prints

\bibitem[{{Delfosse} {et~al.}(2000){Delfosse}, {Forveille}, {S{\'e}gransan},
  {Beuzit}, {Udry}, {Perrier}, \& {Mayor}}]{delfosse2000}
{Delfosse}, X., {Forveille}, T., {S{\'e}gransan}, D., {Beuzit}, J., {Udry}, S.,
  {Perrier}, C., \& {Mayor}, M. 2000, \aap, 364, 217

\bibitem[{{Delorme} {et~al.}(2011){Delorme}, {Collier Cameron}, {Hebb},
  {Rostron}, {Lister}, {Norton}, {Pollacco}, \& {West}}]{delorme2011}
{Delorme}, P., {Collier Cameron}, A., {Hebb}, L., {Rostron}, J., {Lister},
  T.~A., {Norton}, A.~J., {Pollacco}, D., \& {West}, R.~G. 2011, \mnras, 253

\bibitem[{{Dotter} {et~al.}(2008){Dotter}, {Chaboyer}, {Jevremovi{\'c}},
  {Kostov}, {Baron}, \& {Ferguson}}]{dotter2008}
{Dotter}, A., {Chaboyer}, B., {Jevremovi{\'c}}, D., {Kostov}, V., {Baron}, E.,
  \& {Ferguson}, J.~W. 2008, \apjs, 178, 89

\bibitem[{{Efron}(1982)}]{efron1982}
{Efron}, B. 1982, {The Jackknife, the Bootstrap and other resampling plans},
  ed. {Efron, B.}

\bibitem[{{Feigelson} {et~al.}(2004)}]{feigelson04}
{Feigelson}, E.~D., {et~al.} 2004, \apj, 611, 1107

\bibitem[{{Frescura} {et~al.}(2008){Frescura}, {Engelbrecht}, \&
  {Frank}}]{frescura2008}
{Frescura}, F.~A.~M., {Engelbrecht}, C.~A., \& {Frank}, B.~S. 2008, \mnras,
  388, 1693

\bibitem[{{Giampapa} {et~al.}(2006){Giampapa}, {Hall}, {Radick}, \&
  {Baliunas}}]{giampapa2006}
{Giampapa}, M.~S., {Hall}, J.~C., {Radick}, R.~R., \& {Baliunas}, S.~L. 2006,
  \apj, 651, 444

\bibitem[{{Hartman} {et~al.}(2010){Hartman}, {Bakos}, {Kov{\'a}cs}, \&
  {Noyes}}]{hartman2010}
{Hartman}, J.~D., {Bakos}, G.~{\'A}., {Kov{\'a}cs}, G., \& {Noyes}, R.~W. 2010,
  \mnras, 408, 475

\bibitem[{{Henry} \& {McCarthy}(1993)}]{henry1993}
{Henry}, T.~J., \& {McCarthy}, Jr., D.~W. 1993, \aj, 106, 773

\bibitem[{{Hodgkin} {et~al.}(1999){Hodgkin}, {Pinfield}, {Jameson}, {Steele},
  {Cossburn}, \& {Hambly}}]{hodgkin1999}
{Hodgkin}, S.~T., {Pinfield}, D.~J., {Jameson}, R.~F., {Steele}, I.~A.,
  {Cossburn}, M.~R., \& {Hambly}, N.~C. 1999, \mnras, 310, 87

\bibitem[{{Irwin} {et~al.}(2011){Irwin}, {Berta}, {Burke}, {Charbonneau},
  {Nutzman}, {West}, \& {Falco}}]{irwin2011}
{Irwin}, J., {Berta}, Z.~K., {Burke}, C.~J., {Charbonneau}, D., {Nutzman}, P.,
  {West}, A.~A., \& {Falco}, E.~E. 2011, \apj, 727, 56

\bibitem[{{Irwin} \& {Bouvier}(2009)}]{irwin2009}
{Irwin}, J., \& {Bouvier}, J. 2009, in IAU Symposium, Vol. 258, IAU Symposium,
  ed. {E.~E.~Mamajek, D.~R.~Soderblom, \& R.~F.~G.~Wyse}, 363--374

\bibitem[{{Irwin} {et~al.}(2007){Irwin}, {Hodgkin}, {Aigrain}, {Hebb},
  {Bouvier}, {Clarke}, {Moraux}, \& {Bramich}}]{irwin2007}
{Irwin}, J., {Hodgkin}, S., {Aigrain}, S., {Hebb}, L., {Bouvier}, J., {Clarke},
  C., {Moraux}, E., \& {Bramich}, D.~M. 2007, \mnras, 377, 741

\bibitem[{{Jones} {et~al.}(1996){Jones}, {Fischer}, \& {Stauffer}}]{jones1996}
{Jones}, B.~F., {Fischer}, D.~A., \& {Stauffer}, J.~R. 1996, \aj, 112, 1562

\bibitem[{{Kawaler}(1988)}]{kawaler1988}
{Kawaler}, S.~D. 1988, \apj, 333, 236

\bibitem[{{Kiraga} \& {St\k{e}pie{\'n}}(2007)}]{kiraga2007}
{Kiraga}, M., \& {St\k{e}pie{\'n}}, K. 2007, Acta Astronomica, 57, 149

\bibitem[{{Kraus} \& {Hillenbrand}(2007)}]{adam2007}
{Kraus}, A.~L., \& {Hillenbrand}, L.~A. 2007, \aj, 134, 2340

\bibitem[{{Krishnamurthi} {et~al.}(1997){Krishnamurthi}, {Pinsonneault},
  {Barnes}, \& {Sofia}}]{krish1997}
{Krishnamurthi}, A., {Pinsonneault}, M.~H., {Barnes}, S., \& {Sofia}, S. 1997,
  \apj, 480, 303

\bibitem[{{Law} {et~al.}(2010){Law}, {Dekany}, {Rahmer}, {Hale}, {Smith},
  {Quimby}, {Ofek}, {Kasliwal}, {Zolkower}, {Velur}, {Henning}, {Bui},
  {McKenna}, {Nugent}, {Jacobsen}, {Walters}, {Bloom}, {Surace}, {Grillmair},
  {Laher}, {Mattingly}, \& {Kulkarni}}]{nick2010}
{Law}, N.~M. {et~al.} 2010, 7735

\bibitem[{{Law} {et~al.}(2011){Law}, {Kraus}, {Street}, {Lister}, {Shporer}, \&
  {Hillenbrand}}]{law2011}
{Law}, N.~M., {Kraus}, A.~L., {Street}, R.~R., {Lister}, T., {Shporer}, A., \&
  {Hillenbrand}, L.~A. 2011, ArXiv e-prints

\bibitem[{{Law} {et~al.}(2009){Law}, {Kulkarni}, {Dekany}, {Ofek}, {Quimby},
  {Nugent}, {Surace}, {Grillmair}, {Bloom}, {Kasliwal}, {Bildsten}, {Brown},
  {Cenko}, {Ciardi}, {Croner}, {Djorgovski}, {van Eyken}, {Filippenko}, {Fox},
  {Gal-Yam}, {Hale}, {Hamam}, {Helou}, {Henning}, {Howell}, {Jacobsen},
  {Laher}, {Mattingly}, {McKenna}, {Pickles}, {Poznanski}, {Rahmer}, {Rau},
  {Rosing}, {Shara}, {Smith}, {Starr}, {Sullivan}, {Velur}, {Walters}, \&
  {Zolkower}}]{nick2009}
{Law}, N.~M. {et~al.} 2009, \pasp, 121, 1395

\bibitem[{{Mamajek} \& {Hillenbrand}(2008)}]{mamajek2008}
{Mamajek}, E.~E., \& {Hillenbrand}, L.~A. 2008, \apj, 687, 1264

\bibitem[{{Meibom} {et~al.}(2011){Meibom}, {Barnes}, {Latham}, {Batalha},
  {Borucki}, {Koch}, {Basri}, {Walkowicz}, {Janes}, {Jenkins}, {Van Cleve},
  {Haas}, {Bryson}, {Dupree}, {Furesz}, {Szentgyorgyi}, {Buchhave}, {Clarke},
  {Twicken}, \& {Quintana}}]{meibom2011b}
{Meibom}, S. {et~al.} 2011, ArXiv e-prints

\bibitem[{{Meibom} {et~al.}(2009){Meibom}, {Mathieu}, \&
  {Stassun}}]{meibom2009}
{Meibom}, S., {Mathieu}, R.~D., \& {Stassun}, K.~G. 2009, \apj, 695, 679

\bibitem[{{Messina} {et~al.}(2010){Messina}, {Parihar}, {Koo}, {Kim}, {Rey}, \&
  {Lee}}]{messina2010}
{Messina}, S., {Parihar}, P., {Koo}, J., {Kim}, S., {Rey}, S., \& {Lee}, C.
  2010, \aap, 513, A29+

\bibitem[{{Monet} {et~al.}(2003){Monet}, {Levine}, {Canzian}, {Ables}, {Bird},
  {Dahn}, {Guetter}, {Harris}, {Henden}, {Leggett}, {Levison}, {Luginbuhl},
  {Martini}, {Monet}, {Munn}, {Pier}, {Rhodes}, {Riepe}, {Sell}, {Stone},
  {Vrba}, {Walker}, {Westerhout}, {Brucato}, {Reid}, {Schoening}, {Hartley},
  {Read}, \& {Tritton}}]{monet2003}
{Monet}, D.~G. {et~al.} 2003, \aj, 125, 984

\bibitem[{{Ochsenbein} {et~al.}(2000){Ochsenbein}, {Bauer}, \&
  {Marcout}}]{Ochsenbein2000}
{Ochsenbein}, F., {Bauer}, P., \& {Marcout}, J. 2000, \aaps, 143, 23

\bibitem[{{Pace} \& {Pasquini}(2004)}]{pace2004}
{Pace}, G., \& {Pasquini}, L. 2004, \aap, 426, 1021

\bibitem[{{Radick} {et~al.}(1987){Radick}, {Thompson}, {Lockwood}, {Duncan}, \&
  {Baggett}}]{radick1987}
{Radick}, R.~R., {Thompson}, D.~T., {Lockwood}, G.~W., {Duncan}, D.~K., \&
  {Baggett}, W.~E. 1987, \apj, 321, 459

\bibitem[{{Rahmer} {et~al.}(2008){Rahmer}, {Smith}, {Velur}, {Hale}, {Law},
  {Bui}, {Petrie}, \& {Dekany}}]{rahmer2008}
{Rahmer}, G., {Smith}, R., {Velur}, V., {Hale}, D., {Law}, N., {Bui}, K.,
  {Petrie}, H., \& {Dekany}, R. 2008, in Society of Photo-Optical
  Instrumentation Engineers (SPIE) Conference Series, Vol. 7014, Society of
  Photo-Optical Instrumentation Engineers (SPIE) Conference Series

\bibitem[{{Rau} {et~al.}(2009){Rau}, {Kulkarni}, {Law}, {Bloom}, {Ciardi},
  {Djorgovski}, {Fox}, {Gal-Yam}, {Grillmair}, {Kasliwal}, {Nugent}, {Ofek},
  {Quimby}, {Reach}, {Shara}, {Bildsten}, {Cenko}, {Drake}, {Filippenko},
  {Helfand}, {Helou}, {Howell}, {Poznanski}, \& {Sullivan}}]{rau2009}
{Rau}, A. {et~al.} 2009, \pasp, 121, 1334

\bibitem[{{Scholz} \& {Eisl{\"o}ffel}(2007)}]{scholz2007}
{Scholz}, A., \& {Eisl{\"o}ffel}, J. 2007, \mnras, 381, 1638

\bibitem[{{Scholz} {et~al.}(2011){Scholz}, {Irwin}, {Bouvier}, {Sip{\H o}cz},
  {Hodgkin}, \& {Eisl{\"o}ffel}}]{scholz2011}
{Scholz}, A., {Irwin}, J., {Bouvier}, J., {Sip{\H o}cz}, B.~M., {Hodgkin}, S.,
  \& {Eisl{\"o}ffel}, J. 2011, \mnras, 261

\bibitem[{{Skrutskie} {et~al.}(2006){Skrutskie}, {Cutri}, {Stiening},
  {Weinberg}, {Schneider}, {Carpenter}, {Beichman}, {Capps}, {Chester},
  {Elias}, {Huchra}, {Liebert}, {Lonsdale}, {Monet}, {Price}, {Seitzer},
  {Jarrett}, {Kirkpatrick}, {Gizis}, {Howard}, {Evans}, {Fowler}, {Fullmer},
  {Hurt}, {Light}, {Kopan}, {Marsh}, {McCallon}, {Tam}, {Van Dyk}, \&
  {Wheelock}}]{2mass}
{Skrutskie}, M.~F. {et~al.} 2006, \aj, 131, 1163

\bibitem[{{Skumanich}(1972)}]{skumanich72}
{Skumanich}, A. 1972, \apj, 171, 565

\bibitem[{{Soderblom} {et~al.}(2001){Soderblom}, {Jones}, \&
  {Fischer}}]{soderblom2001}
{Soderblom}, D.~R., {Jones}, B.~F., \& {Fischer}, D. 2001, \apj, 563, 334

\bibitem[{{Stauffer} {et~al.}(1997){Stauffer}, {Balachandran}, {Krishnamurthi},
  {Pinsonneault}, {Terndrup}, \& {Stern}}]{stauffer1997}
{Stauffer}, J.~R., {Balachandran}, S.~C., {Krishnamurthi}, A., {Pinsonneault},
  M., {Terndrup}, D.~M., \& {Stern}, R.~A. 1997, \apj, 475, 604

\bibitem[{{Steele} \& {Jameson}(1995)}]{steele1995}
{Steele}, I.~A., \& {Jameson}, R.~F. 1995, \mnras, 272, 630

\bibitem[{{Stoughton} {et~al.}(2002){Stoughton}, {Lupton}, {Bernardi},
  {Blanton}, {Burles}, {Castander}, {Connolly}, {Eisenstein}, {Frieman},
  {Hennessy}, {Hindsley}, {Ivezi{\'c}}, {Kent}, {Kunszt}, {Lee}, {Meiksin},
  {Munn}, {Newberg}, {Nichol}, {Nicinski}, {Pier}, {Richards}, {Richmond},
  {Schlegel}, {Smith}, {Strauss}, {SubbaRao}, {Szalay}, {Thakar}, {Tucker},
  {Vanden Berk}, {Yanny}, {Adelman}, {Anderson}, {Anderson}, {Annis},
  {Bahcall}, {Bakken}, {Bartelmann}, {Bastian}, {Bauer}, {Berman},
  {B{\"o}hringer}, {Boroski}, {Bracker}, {Briegel}, {Briggs}, {Brinkmann},
  {Brunner}, {Carey}, {Carr}, {Chen}, {Christian}, {Colestock}, {Crocker},
  {Csabai}, {Czarapata}, {Dalcanton}, {Davidsen}, {Davis}, {Dehnen},
  {Dodelson}, {Doi}, {Dombeck}, {Donahue}, {Ellman}, {Elms}, {Evans}, {Eyer},
  {Fan}, {Federwitz}, {Friedman}, {Fukugita}, {Gal}, {Gillespie}, {Glazebrook},
  {Gray}, {Grebel}, {Greenawalt}, {Greene}, {Gunn}, {de Haas}, {Haiman},
  {Haldeman}, {Hall}, {Hamabe}, {Hansen}, {Harris}, {Harris}, {Harvanek},
  {Hawley}, {Hayes}, {Heckman}, {Helmi}, {Henden}, {Hogan}, {Hogg}, {Holmgren},
  {Holtzman}, {Huang}, {Hull}, {Ichikawa}, {Ichikawa}, {Johnston}, {Kauffmann},
  {Kim}, {Kimball}, {Kinney}, {Klaene}, {Kleinman}, {Klypin}, {Knapp},
  {Korienek}, {Krolik}, {Kron}, {Krzesi{\'n}ski}, {Lamb}, {Leger},
  {Limmongkol}, {Lindenmeyer}, {Long}, {Loomis}, {Loveday}, {MacKinnon},
  {Mannery}, {Mantsch}, {Margon}, {McGehee}, {McKay}, {McLean}, {Menou},
  {Merelli}, {Mo}, {Monet}, {Nakamura}, {Narayanan}, {Nash}, {Neilsen},
  {Newman}, {Nitta}, {Odenkirchen}, {Okada}, {Okamura}, {Ostriker}, {Owen},
  {Pauls}, {Peoples}, {Peterson}, {Petravick}, {Pope}, {Pordes}, {Postman},
  {Prosapio}, {Quinn}, {Rechenmacher}, {Rivetta}, {Rix}, {Rockosi}, {Rosner},
  {Ruthmansdorfer}, {Sandford}, {Schneider}, {Scranton}, {Sekiguchi}, {Sergey},
  {Sheth}, {Shimasaku}, {Smee}, {Snedden}, {Stebbins}, {Stubbs}, {Szapudi},
  {Szkody}, {Szokoly}, {Tabachnik}, {Tsvetanov}, {Uomoto}, {Vogeley}, {Voges},
  {Waddell}, {Walterbos}, {Wang}, {Watanabe}, {Weinberg}, {White}, {White},
  {Wilhite}, {Wolfe}, {Yasuda}, {York}, {Zehavi}, \& {Zheng}}]{stoughton02}
{Stoughton}, C. {et~al.} 2002, \aj, 123, 485

\bibitem[{{Terndrup} {et~al.}(2000){Terndrup}, {Stauffer}, {Pinsonneault},
  {Sills}, {Yuan}, {Jones}, {Fischer}, \& {Krishnamurthi}}]{terndrup2000}
{Terndrup}, D.~M., {Stauffer}, J.~R., {Pinsonneault}, M.~H., {Sills}, A.,
  {Yuan}, Y., {Jones}, B.~F., {Fischer}, D., \& {Krishnamurthi}, A. 2000, \aj,
  119, 1303

\bibitem[{{van Leeuwen}(2009)}]{vanleeuwen2009}
{van Leeuwen}, F. 2009, \aap, 497, 209

\bibitem[{{West} {et~al.}(2008){West}, {Hawley}, {Bochanski}, {Covey}, {Reid},
  {Dhital}, {Hilton}, \& {Masuda}}]{andy08}
{West}, A.~A., {Hawley}, S.~L., {Bochanski}, J.~J., {Covey}, K.~R., {Reid},
  I.~N., {Dhital}, S., {Hilton}, E.~J., \& {Masuda}, M. 2008, \aj, 135, 785

\bibitem[{{York} {et~al.}(2000)}]{york00}
{York}, D.~G., {et~al.} 2000, \aj, 120, 1579

\end{thebibliography}
\end{document}